\documentclass[10pt,twocolumn,twoside]{IEEEtran} 
\ifCLASSINFOpdf
   \usepackage[pdftex]{graphicx}
\else
   \usepackage[dvips]{graphicx}
\fi
\usepackage[cmex10]{amsmath}
\usepackage{amsfonts,amssymb}
\usepackage{algorithm}
\usepackage{algorithmicx}
\usepackage{algpseudocode}
\usepackage{color}

\newcommand{\eq}{\triangleq}
\DeclareMathOperator*{\argmin}{arg\,min}
\DeclareMathOperator*{\supp}{supp}
\DeclareMathOperator{\sgn}{sgn}
\DeclareMathOperator{\sat}{sat}

\newcommand{\field}[1]{\mathbb{#1}}
\newcommand{\R}{\field{R}}

\newcommand{\U}{{\mathcal{U}}}
\newcommand{\shrink}{S}
\newcommand{\spr}{R}
\newcommand{\meas}{m_{L}}

\newcommand{\vc}[1]{{\boldsymbol{#1}}}
\newcommand{\dez}{D}

\newtheorem{thm}{Theorem}

\newtheorem{lem}[thm]{Lemma}
\newtheorem{prop}[thm]{Proposition}
\newtheorem{defn}[thm]{Definition}
\newtheorem{rem}[thm]{Remark}
\newtheorem{problem}[thm]{Problem}
\newtheorem{example}[thm]{Example}

\begin{document}
%
\title{
Maximum Hands-Off Control: A Paradigm of Control Effort Minimization
\thanks{A preliminary version of parts of this work was presented in \cite{NagQueNes13}.}
}
%
%
%
\author{
 Masaaki Nagahara,~\IEEEmembership{Senior Member,~IEEE,}\\
 Daniel E. Quevedo,~\IEEEmembership{Senior Member,~IEEE,}\\
 Dragan Ne\v{s}i\'{c},~\IEEEmembership{Fellow,~IEEE}%
\thanks{
This research is supported in part by the JSPS Grant-in-Aid for Scientific Research (C) No.~24560543,
MEXT Grant-in-Aid for Scientific  Research on Innovative Areas
No.~26120521, and an Okawa Foundation Research Grant. 
}%
\thanks{
M. Nagahara is with
 Graduate School of Informatics, Kyoto
 University, Kyoto, 606-8501, 
 Japan; 
 email: nagahara@ieee.org}%
\thanks{
D. E. Quevedo is with School of Electrical Engineering \& 
 Computer Science, The University of Newcastle, NSW
 2308, Australia;
 email: dquevedo@ieee.org}%
\thanks{D. Ne\v{s}i\'{c} is with
 Department of Electrical and Electronic Engineering, 
 The University of Melbourne, Victoria 3010
 Australia; email: dnesic@unimelb.edu.au}
}

%
%

\markboth{Journal of \LaTeX\ Class Files,~Vol.~11, No.~4, December~2012}%
{Shell \MakeLowercase{\textit{et al.}}: Bare Demo of IEEEtran.cls for Journals}
%



\maketitle

\begin{abstract}
In this paper, 
we propose a paradigm of control,
called a maximum hands-off control.
A hands-off control is defined as a control that has
a short support per unit time.
The maximum hands-off control is the minimum support (or sparsest) per unit time
among all controls that achieve control objectives.
For finite horizon continuous-time control,
we show the equivalence between the maximum hands-off control and 
$L^1$-optimal control 
under a uniqueness assumption called normality.
This result rationalizes the use of $L^1$ optimality in computing a maximum hands-off control.
The same result is obtained for discrete-time hands-off control.
We also propose an $L^1$/$L^2$-optimal control to obtain a smooth hands-off control.
Furthermore, we give a self-triggered feedback control algorithm for linear time-invariant systems,
which achieves a given sparsity rate
and practical stability in the case of plant disturbances.
An example is included to illustrate the effectiveness of the proposed control.
\end{abstract}


%
\IEEEpeerreviewmaketitle

\section{Introduction}
\label{sec:introduction}
%
%
%
%
In practical control systems, 
we often need to
minimize the control effort
so as to achieve control objectives
under limitations in equipment such as
actuators, sensors, and networks.
For example, 
the energy (or $L^2$-norm) of a control signal can be minimized
to prevent engine overheating
or to reduce transmission cost
by means of a standard LQ (linear quadratic) control problem;
see e.g., \cite{AndMoo}.
Another example is the \emph{minimum fuel} control,
discussed in e.g., \cite{Ath63,AthFal},
in which the total expenditure of fuel is minimized with
the $L^1$ norm of the control.

Alternatively, in some situations, the control effort can be dramatically reduced by
holding the control value \emph{exactly zero} over a time interval.
We call such control a \emph{hands-off control}.
A motivation for hands-off control is a stop-start system
in automobiles.
It is a hands-off control; it automatically shuts down 
the engine to avoid it idling for long periods of time.
By this, we can reduce CO or CO2 emissions as well as fuel consumption
\cite{Dun74}.
This strategy is also used 
in electric/hybrid vehicles \cite{Cha07};
the internal combustion engine is stopped when
the vehicle is at a stop or the speed is lower than a preset threshold,
and the electric motor is alternatively used.
Thus hands-off control also has potential for solving environmental problems.
In railway vehicles,
hands-off control, called \emph{coasting}, is used to reduce energy consumption
\cite{LiuGol03}.
Furthermore, hands-off control is
desirable for networked and embedded systems
since the communication channel is not used
during a period of zero-valued control.
This property is advantageous in particular for wireless communications
\cite{JeoJeo06,KonWonTsa09}
and networked control systems
\cite{NagQue11,HuaShi12,NagQueOst14,KonGooSer14}.
Motivated by these applications, we propose 
a paradigm of control, called \emph{maximum hands-off control}
that maximizes the time interval over which the control is exactly zero.
%

The hands-off property is related to \emph{sparsity},
or the \emph{$L^0$ ``norm''} 
(the quotation marks indicate that this is not a norm;
see Section~\ref{sec:preliminaries} below)
of a signal,
defined by the total length of the intervals over which the signal
takes non-zero values. The maximum hands-off control,
in other words, seeks 
the \emph{sparsest} (or $L^0$-optimal) control among all admissible controls.
The notion of sparsity has been recently adapted to control systems,
including works on model predictive control \cite{NagQue11,Gis+13,HarGalMac13,PakOhlLju13,NagQueOst14},
system gain analysis \cite{SchEbeAll11},
sparse controller design \cite{FarLinJov11},
state estimation \cite{ChaAsiRomRoz11},
to name a few.
The maximum hands-off control is also related to the
minimum attention control \cite{Bro97},
and also to the approach by Donkers et al. \cite{DonTabHee14},
which maximizes the time between consecutive execution of the 
control tasks.
The minimum attention control minimizes
the number of switching per unit time.
In contrast, the maximum hands-off control does not
necessarily minimize the number of switching,
although we show this number is bounded
for linear systems.

The maximum hands-off control (or $L^0$-optimal control) problem
is hard to solve since the cost function
is non-convex and discontinuous.%
\footnote{Very recently, $L^p$ control with $p\in[0,1)$ has
been investigated in \cite{ItoKun14},
which introduces regularization terms to guarantee the
existence of optimal solutions.}
To overcome the difficulty, one can adopt
$L^1$ optimality as a convex relaxation of the problem,
as often used in \emph{compressed sensing}
\cite{Don06,Can06}.
Compressed sensing has shown by theory and experiments that 
sparse high-dimensional signals
can be reconstructed from incomplete measurements
by using $\ell^1$ optimization;
see e.g., \cite{Ela,EldKut,HayNagTan13} for details.

Interestingly,  a finite horizon $L^1$-optimal (or minimum fuel) control has been known to have
such a sparsity property, traditionally called \emph{"bang-off-bang"}
\cite{AthFal}.
Based on this, $L^1$-optimal control has been recently investigated for designing sparse control
\cite{KosKosFed10,CasHerWac12,JovLin13}.
Although advantage has implicitly been taken of the sparsity property for minimizing the $L^1$ norm,
we are not aware of results
on the theoretical connection between sparsity and
$L^1$ optimality of the control.
In the present manuscript, we prove that 
a solution to an $L^1$-optimal control problem gives a maximum hands-off control, and vice versa.
As a result,
the sparsest solution (i.e., the maximum hands-off control)
can be obtained by solving an $L^1$-optimal control problem.
The same result is obtained for discrete-time hands-off control.
We also propose $L^1$/$L^2$-optimal control to avoid the discontinuous
property of "bang-off-bang" in maximum hands-off control.
We show that the $L^1$/$L^2$-optimal control is an intermediate control
between the maximum hands-off (or $L^1$-optimal) control
and the minimum energy (or $L^2$-optimal) control,
in the sense that the $L^1$ and $L^2$ controls are
the limiting instances
of the $L^1$/$L^2$-optimal control.

We also extend the maximum hands-off control to feedback control
for linear time-invariant, reachable, and nonsingular systems
by a \emph{self-triggering} approach
\cite{WanLem09,MazAntTab10,HenQueSanJoh12,BerGomHee12}.
For this, we define sparsity of infinite horizon control signals by
the \emph{sparsity rate}, the $L^0$ norm per unit time.
We give a self-triggered feedback control algorithm that achieves a given sparsity rate
and practical stability in the presence of plant disturbances.
Simulations studies demonstrate the effectiveness of the proposed control method.

The present manuscript expands upon our recent conference contribution \cite{NagQueNes13}
by incorporating feedback control into the formulation.

The remainder of this article is organized as follows:
In Section~\ref{sec:preliminaries},
we give mathematical preliminaries for our subsequent discussion.
In Section~\ref{sec:MHOC-problem},  we formulate the 
maximum hands-off control problem.
Section~\ref{sec:MHOC-solution} is the main part of this paper, 
in which we introduce $L^1$-optimal control as relaxation of the maximum hands-off control,
and establish the theoretical connection between them.
We also analyze discrete-time hands-off control in this section.
In Section~\ref{sec:L1L2},
we propose $L^1$/$L^2$-optimal control for a smooth hands-off control in this section.
In Section~\ref{sec:feedback}, we address the feedback hands-off control.
Section~\ref{sec:examples} presents control design examples
to illustrate the effectiveness of our method.
In Section~\ref{sec:conclusion}, we offer concluding remarks.

\section{Mathematical Preliminaries}
\label{sec:preliminaries}
For a vector $\vc{x}\in\R^n$, we define its norm by
\[
 \|\vc{x}\| \eq \sqrt{\vc{x}^\top\vc{x}},
\]
and for a matrix $A\in\R^{n\times n}$,
\[
 \|A\| \eq \max_{\vc{x}\in\R^n, \|\vc{x}\|=1} \|A\vc{x}\|.
\]  
For a continuous-time signal $u(t)$
over a time interval $[0,T]$,
we define its $L^p$ norm with $p\in[1,\infty)$
by
\begin{equation}
 \|u\|_{p} \eq \left(\int_0^T |u(t)|^p dt\right)^{1/p},
 \label{eq:p-norm}
\end{equation}
and let $L^p[0,T]$ consist of all $u$ for which $\|u\|_p<\infty$.
Note that we can also define \eqref{eq:p-norm} for $p\in(0,1)$, which is not a
norm (It fails to satisfy the triangle inequality.).
We define the support set of $u$, denoted by $\supp(u)$,
the closure of the set
\[
\{t\in[0,T]: u(t)\neq0\}.
\]
Then we define
the $L^0$ ``norm'' of measurable function $u$ as
the length of its support, that is,
\[
 \|u\|_{0} \eq \meas\bigl(\supp(u)\bigr),
\] 
where $\meas$ is the Lebesgue measure on $\R$.
Note that the $L^0$ ``norm'' is not a norm since
it fails to satisfy the positive homogeneity property, that is,
for any non-zero scalar $\alpha$ such that
$|\alpha|\neq 1$, we have
\[
 \|\alpha u\|_{0} = \|u\|_{0} \neq |\alpha| \|u\|_{0},
 \quad \forall u\neq 0.
\]
The notation $\|\cdot\|_{0}$ may be however
justified from the fact that
if $u\in L^1[0,T]$, then
$\|u\|_p<\infty$ for any $p\in(0,1)$ and
\[
 \lim_{p\to 0}\|u\|_{p}^p = \|u\|_{0},
\]
which can be proved by using Lebesgue's monotone convergence theorem
\cite{Rud}.
For more details of $L^p$ when $p\in[0,1)$, 
see \cite{KalPecRob}.
For a function $\vc{f}=[f_1,\dots,f_n]^\top: \R^n\rightarrow\R^n$,
the Jacobian $\vc{f}'$ is defined by
\[
 \vc{f}'(\vc{x}) \eq 
  \begin{bmatrix}
   \frac{\partial f_1}{\partial x_1}&\dots&\frac{\partial f_1}{\partial x_n}\\
   \vdots & \ddots & \vdots\\
   \frac{\partial f_n}{\partial x_1}&\dots&\frac{\partial f_n}{\partial x_n}\\
 \end{bmatrix},
\]  
where $\vc{x}=[x_1,\dots,x_n]^\top$.
For functions $f$ and $g$, 
we denote by $f\circ g$ the composite function
$f(g(\cdot))$.

\section{Maximum Hands-Off Control Problem}
\label{sec:MHOC-problem}
In this section, we formulate the maximum hands-off control problem.
We first define the \emph{sparsity rate},
the $L^0$ norm of a signal per unit time, of finite-horizon
continuous-time signals.
\begin{defn}[Sparsity rate]
\label{defn:sparsity-rate-f}
For measurable function $u$ on $[0,T]$, $T>0$,
the \emph{sparsity rate} is defined by
\begin{equation}
 \spr_T(u) := \frac{1}{T} \|u\|_{0}.
 \label{eq:sparsity_rate_f}
\end{equation}
\end{defn}
Note that for any measurable $u$, $0\leq \spr_T(u)\leq 1$.
If $\spr_T(u)\ll 1$, we say $u$ is \emph{sparse}.%
\footnote{This is analogous to the sparsity of a vector.
When a vector has a small number of non-zero elements relative to the vector size,
then it is called \emph{sparse}.
See \cite{Ela,EldKut,HayNagTan13} for details.}
The control objective is, roughly speaking, 
to design a control $u$ which is as sparse as possible,
whilst satisfying performance criteria. 
For that purpose, we will first focus on finite $T$ and then, 
in Section~\ref{sec:feedback}, study the infinite horizon case, where $T\to\infty$.

To formulate the control problem, we consider nonlinear multi-input plant models of the form
\begin{equation}
 \frac{d\vc{x}(t)}{dt} = \vc{f}\bigl(\vc{x}(t)\bigr) + \sum_{i=1}^m \vc{g}_i\bigl(\vc{x}(t)\bigr)u_i(t),
 \quad t\in[0,T],
 \label{eq:plant}
\end{equation}
where
$\vc{x}(t)\in{\mathbb{R}}^n$ is the state,
$u_1,\dots,u_m$ are the scalar control inputs,
$\vc{f}$ and $\vc{g}_i$
are functions on $\R^n$.
We assume that $\vc{f}(\vc{x})$, $\vc{g}_i(\vc{x})$,
and their Jacobians $\vc{f}'(\vc{x})$, $\vc{g}_i'(\vc{x})$
are continuous.
We use the vector representation $\vc{u}\eq[u_1,\dots,u_m]^\top$.

The control $\{\vc{u}(t): t\in[0,T]\}$ is chosen to drive the state $\vc{x}(t)$
from a given initial state 
\begin{equation}
 \vc{x}(0)=\vc{\xi},
 \label{eq:initial_state}
\end{equation} 
to the origin at a fixed final time $T>0$, that is,
\begin{equation}
 \vc{x}(T)=\vc{0}.
 \label{eq:final_state}
\end{equation}
Also, the components of the control $\vc{u}(t)$ are constrained in magnitude by
\begin{equation}
 \max_i |u_i(t)| \leq 1,
 \label{eq:input_constraint}
\end{equation}
for all $t\in [0,T]$.
We call a control $\{\vc{u}(t): t\in[0,T]\}\in L^1[0,T]$ \emph{admissible}
if it satisfies \eqref{eq:input_constraint} for all $t\in [0,T]$,
and the resultant state $\vc{x}(t)$ from \eqref{eq:plant} satisfies boundary conditions
\eqref{eq:initial_state} and \eqref{eq:final_state}.
We denote by $\U(T,\vc{\xi})$ the set of all admissible controls.

To consider control in $\U(T,\vc{\xi})$, it is necessary that
$\U(T,\vc{\xi})$ is non empty.
This property is basically related to the
\emph{minimum-time control}
formulated as follows:
\begin{problem}[Minimum-time control]
\label{prob:minimum-time}
Find a control $\vc{u}\in L^1[0,T]$ that satisfies \eqref{eq:input_constraint},
and drives $\vc{x}$ from initial state $\vc{\xi}\in\R^n$ to the origin $\vc{0}$
in minimum time.
\hfill\IEEEQEDhere
\end{problem}

Let $T^\ast(\vc{\xi})$ denote the minimum time (or the value function) of
Problem~\ref{prob:minimum-time}.
Also, we define the reachable set as follows:%
\footnote{For linear systems, the reachable set is known to have
nice properties such as convexity and compactness \cite{HerLas,Haj71}.}
\begin{defn}[Reachable set]
\label{defn:reachable-set}
We define the reachable set at time $t\in [0,\infty)$ by
\begin{equation}
 {\mathcal R}(t) \eq \left\{\vc{\xi}\in\R^n: T^\ast(\vc{\xi})\leq t\right\}.
 \label{eq:reachable-set-t}
\end{equation}
and the reachability set
\begin{equation}
 {\mathcal R} \eq \bigcup_{t\geq 0} {\mathcal R}(t).
 \label{eq:reachable-set}
\end{equation}
\end{defn}
To guarantee that $\U(T,\vc{\xi})$ is non-empty,
we introduce the standing assumptions:
\begin{enumerate}
\item $\vc{\xi}\in {\mathcal R}$,
\item $T > T^\ast(\vc{\xi})$.
\end{enumerate}

Now let us formulate our control problem. 
The \emph{maximum hands-off control}
is a control that is the \emph{sparsest}
among all admissible controls in $\U(T,\vc{\xi})$.
In other words, we try to find a control that
maximizes the time interval over which the control $\vc{u}(t)$ is exactly zero.
\footnote{
More precisely, the maximum hands-off control 
minimizes
the Lebesgue measure
of the support. Hence, the values on the sets of measure zero are ignored
and treated as zero in this setup.}
We state the associated optimal control problem as follows:
\begin{problem}[Maximum hands-off control]
\label{prob:MHOC}
Find an admissible control on $[0,T]$, $\vc{u}\in\U(T,\vc{\xi})$,  that minimizes
the sum of sparsity rates:
\begin{equation}
 J_0(\vc{u}) \eq \sum_{i=1}^m\lambda_i \spr_T(u_i) = \frac{1}{T}\sum_{i=1}^m\lambda_i \|u_i\|_{0},
 \label{eq:J_MHO}
\end{equation}
where $\lambda_1>0,\dots,\lambda_m>0$ are given weights.
\hfill\IEEEQEDhere
\end{problem}

This control problem is quite difficult to solve since the objective function is highly nonlinear and non-smooth.
In the next section, we discuss convex relaxation of the maximum hands-off control problem,
which gives the exact solution of Problem~\ref{prob:MHOC} under some assumptions.

\begin{rem}
The input constraint \eqref{eq:input_constraint} is necessary.
Let us consider the integrator $\dot{x}(t)=u(t)$ and 
remove the constraint $\eqref{eq:input_constraint}$.
Then for any $\epsilon>0$, the following control is an admissible control
\[
 u_\epsilon(t) = \begin{cases} \xi/\epsilon, & t \in [0,\epsilon),\\ 0, & t\in [\epsilon,T],\end{cases}
\]
which has arbitrarily small $L^0$ norm.
But $\lim_{\epsilon\to 0}u_\epsilon$ is not a function,
so called Dirac's delta, and hence is not in $L^1$.
In this case, the maximum hands-off problem has no solution.
\end{rem}

\section{Solution to Maximum Hands-Off Control Problem}
\label{sec:MHOC-solution}
In this section we will show how the maximum hands-off control can be solved in closed form.
\subsection{Convex Relaxation}
\label{subsec:convex-relaxation}
Here we consider convex relaxation of the maximum hands-off control problem.
We replace $\|u_i\|_{0}$ in \eqref{eq:J_MHO}
with $L^1$ norm $\|u_i\|_{1}$,
and obtain the following \emph{$L^1$-optimal control} problem,
also known as \emph{minimum fuel control}
discussed in e.g. \cite{Ath63,AthFal}.
\begin{problem}[$L^1$-optimal control]
\label{prob:L1}
Find an admissible control $\vc{u}\in\U(T,\vc{\xi})$ on $[0,T]$ that minimizes
\begin{equation}
 J_1(\vc{u}) \eq \frac{1}{T}\sum_{i=1}^m \lambda_i \|u_i\|_{1} 
  = \frac{1}{T} \sum_{i=1}^m \lambda_i \int_0^T |u_i(t)| dt,
 \label{eq:J_L1}
\end{equation}
where $\lambda_1>0,\dots,\lambda_m>0$ are given weights.
\hfill\IEEEQEDhere
\end{problem}

The objective function \eqref{eq:J_L1} is convex in $\vc{u}$ and this control problem
is much easier to solve than the maximum hands-off control problem (Problem~\ref{prob:MHOC}).
The main contribution of this section is that we prove the solution set of Problem~\ref{prob:L1} 
is equivalent to that of Problem~\ref{prob:MHOC},
under the assumption of normality.
Before proving this property, we review $L^1$-optimal control in the next subsection.


\subsection{Review of $L^1$-Optimal Control}
\label{subsec:L1}
Here we briefly review the $L^1$-optimal control problem (Problem~\ref{prob:L1})
based on the discussion in \cite[Section 6-13]{AthFal}.

\begin{figure}[tb]
\centering
\includegraphics[width=\linewidth]{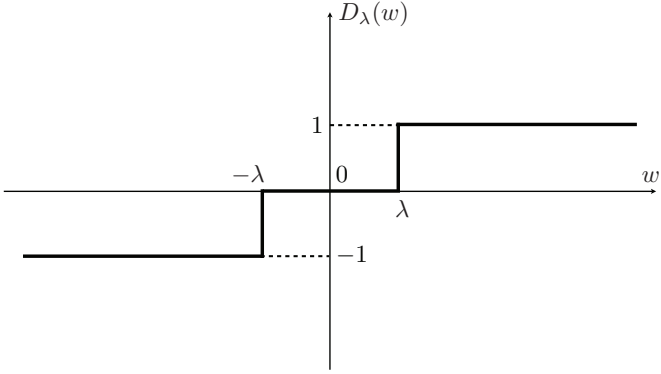}
\caption{Dead-zone function $\dez_\lambda(w)$}
\label{fig:dez}
\end{figure}

Let us first form the Hamiltonian function for the $L^1$-optimal control problem as
\begin{equation}
 H(\vc{x},\vc{p},\vc{u}) = \frac{1}{T}\sum_{i=1}^m \lambda_i |u_i| 
  + \vc{p}^\top \biggl(\vc{f}\bigl(\vc{x}\bigr)
    +\sum_{i=1}^m \vc{g}_i(\vc{x})u_i\biggr),
 \label{eq:Hamiltonian_L1}
\end{equation}
where 
$\vc{p}$ is the costate (or adjoint) vector
\cite[Section 5-7]{AthFal}.
Assume that $\vc{u}^\ast=[u_1^\ast,\dots,u_m^\ast]^\top$
is an $L^1$-optimal control
and $\vc{x}^\ast$ is the resultant state trajectory.
According to the minimum principle, there exists a costate $\vc{p}^\ast$ such that
the optimal control $\vc{u}^\ast$ satisfies
\[
 H\bigl(\vc{x}^\ast(t),\vc{p}^\ast(t),\vc{u}^\ast(t)\bigr)\leq  H\bigl(\vc{x}^\ast(t),\vc{p}^\ast(t),\vc{u}(t)\bigr),
\]
for for all $t\in[0,T]$ and all admissible $\vc{u}$.
The optimal state $\vc{x}^\ast$ and costate $\vc{p}^\ast$ 
satisfies the canonical equations
\[
 \begin{split}
  \frac{d\vc{x}^\ast(t)}{dt} &=  \vc{f}\bigl(\vc{x}^\ast(t)\bigr) + \sum_{i=1}^m
    \vc{g}_i\bigl(\vc{x}^\ast(t)\bigr)u_i^\ast(t),\\
  \frac{d\vc{p}^\ast(t)}{dt} &= - \vc{f}'\bigl(\vc{x}^\ast(t)\bigr)^\top \vc{p}^\ast(t)\\
  	&\qquad-\sum_{i=1}^m u_i^\ast(t) \vc{g}_i'\bigl(\vc{x}^\ast(t)\bigr)^\top \vc{p}^\ast(t),
 \end{split} 
\]
with boundary conditions
\[
 \vc{x}^\ast(0) = \vc{\xi},\quad \vc{x}^\ast(T) = \vc{0}.
\]
The minimizer
$\vc{u}^\ast=[u_1^\ast,\dots,u_m^\ast]^\top$ of the Hamiltonian in 
\eqref{eq:Hamiltonian_L1} is given by
\[
 u_i^\ast(t) = -\dez_{\lambda_i/T}\left(\vc{g}_i\bigl(\vc{x}^\ast(t)\bigr)^\top \vc{p}^\ast(t)\right),\quad t\in[0,T],
\] 
where $\dez_{\lambda}(\cdot):\R^n\rightarrow[-1,1]$ is the dead-zone
(set-valued)
function defined by
\begin{equation}
 \begin{split}
 \dez_{\lambda}(w)  &=
   \begin{cases} 
     -1,& \text{~if~} w<-\lambda,\\
     0, & \text{~if~} -\lambda<w<\lambda,\\
     1, & \text{~if~} \lambda<w,\\
   \end{cases}\\
  \dez_{\lambda}(w) &\in [-1,0], \text{~if~} w=-\lambda,\\
  \dez_{\lambda}(w) &\in [0,1], \text{~if~} w=\lambda.
 \end{split}
 \label{eq:dez}
\end{equation}
See Fig.~\ref{fig:dez} for the graph of $\dez_\lambda(\cdot)$.

If $\vc{g}_i(\vc{x}^\ast)^\top \vc{p}^\ast$ is equal to $-\lambda_i/T$ or $\lambda_i/T$ over
a non-zero time interval, say $[t_1,t_2]\subset[0,T]$, where
$t_1<t_2$,
then the control $u_i$
(and hence $\vc{u}$)
over $[t_1,t_2]$
cannot be uniquely determined
by the minimum principle.
In this case, the interval $[t_1,t_2]$ is called a \emph{singular interval},
and a control problem that has at least one singular interval is called
\emph{singular}.
If there is no singular interval, the problem is called \emph{normal}:
\begin{defn}[Normality]
The $L^1$-optimal control problem stated in Problem~\ref{prob:L1} is said to be \emph{normal}
if the set 
\[
 {\mathcal T}_i \eq \{t\in [0,T]: |T\lambda_i^{-1}\vc{g}_i(\vc{x}^\ast(t))^\top \vc{p}^\ast(t)|=1\}
\]
is countable for $i=1,\dots,m$.
If the problem is normal, the elements $t_1,t_2,\dots\in {\mathcal T}_i$
are called the \emph{switching times} for the control $u_i(t)$.
\end{defn}

If the problem is normal, the components
of the $L^1$-optimal control $\vc{u}^\ast(t)$
are piecewise constant and ternary,
taking values $\pm 1$ or $0$
at almost all%
\footnote{
Throughout this paper, 
``almost all'' means
``all but a set of Lebesgue measure zero.''
}
$t\in[0,T]$.
This property, named "bang-off-bang,"
is the key to relate the $L^1$-optimal control with
the maximum hands-off control
as discussed in the next section.

In general, it is difficult to check if the problem is normal
without solving the canonical equations
\cite[Section 6-22]{AthFal}.
For linear plants, however, a sufficient condition
for normality is obtained \cite[Theorem 6-13]{AthFal}.
\subsection{Maximum Hands-Off Control and $L^1$ Optimality}
\label{subsec:main}
In this section, we study the relation
between
maximum hands-off control stated in Problem~\ref{prob:MHOC}
and $L^1$-optimal control stated in Problem~\ref{prob:L1}.
The theorem below rationalizes the use of $L^1$ optimality
in computing the maximum hands-off control.
\begin{thm}
\label{thm:L1optimal}
Assume that the $L^1$-optimal control problem (Problem~\ref{prob:L1})
is normal and has at least one solution.
Let $\U_0^\ast$ and $\U_1^\ast$ be the sets of the optimal solutions
of Problem~\ref{prob:MHOC} (maximum hands-off control problem)
and Problem~\ref{prob:L1},
respectively.
Then we have $\U_0^\ast = \U_1^\ast$.
\end{thm}
\begin{IEEEproof}
By assumption, $\U_1^\ast$ is non-empty, 
and so is $\U(T,\vc{\xi})$, the set of all admissible controls.
Also we have $\U_0^\ast \subset \U(T,\vc{\xi})$.
We first show that $\U_0^\ast$ is non-empty, and then prove that $\U_0^\ast = \U_1^\ast$.

First, for any $\vc{u}\in\U(T,\vc{\xi})$, we have
\begin{equation}
 \begin{split}
  J_1(\vc{u}) &= \frac{1}{T}\sum_{i=1}^m \lambda_i \int_0^T |u_i(t)| ~dt\\
  &= \frac{1}{T}\sum_{i=1}^m \lambda_i \int_{\supp(u_i)} |u_i(t)| ~dt\\  
  &\leq \frac{1}{T}\sum_{i=1}^m \lambda_i \int_{\supp(u_i)} 1 ~dt
  = J_0(\vc{u}).
 \end{split} 
\label{eq:proof1}
\end{equation}

Now take an arbitrary $\vc{u}^\ast_1 \in \U_1^\ast$.
Since the problem is normal by assumption,
each control $u_{1i}^\ast(t)$ in $\vc{u}_1^\ast(t)$ takes values $-1$, $0$, or $1$,
at almost all $t\in[0,T]$.
This implies that
\begin{equation}
 \begin{split}
 J_1(\vc{u}^\ast_1) &= \frac{1}{T}\sum_{i=1}^m \lambda_i \int_0^T |u_{1i}^\ast(t)| ~dt\\
  &= \frac{1}{T}\sum_{i=1}^m \lambda_i \int_{\supp(u_{1i}^\ast)} 1 ~dt
  = J_0(\vc{u}^\ast_1).
 \end{split}
 \label{eq:proof2}
\end{equation}
From \eqref{eq:proof1} and \eqref{eq:proof2},
$\vc{u}^\ast_1$ is a minimizer of $J_0$,
that is, $\vc{u}_1^\ast\in\U_0^\ast$.
Thus, $\U_0^\ast$ is non-empty and $\U_1^\ast \subset \U_0^\ast$.

Conversely, let $\vc{u}^\ast_0\in\U_0^\ast\subset\U(T,\vc{\xi})$.
Take independently $\vc{u}^\ast_1\in\U_1^\ast\subset\U(T,\vc{\xi})$.
From \eqref{eq:proof2} and the optimality of $\vc{u}^\ast_1$, we have
\begin{equation}
 J_0(\vc{u}^\ast_1) = J_1(\vc{u}^\ast_1) \leq J_1(\vc{u}^\ast_0).
 \label{eq:proof3}
\end{equation} 
On the other hand, from \eqref{eq:proof1} and the optimality of $\vc{u}^\ast_0$,
we have
\begin{equation}
 J_1(\vc{u}^\ast_0)\leq J_0(\vc{u}^\ast_0) \leq J_0(\vc{u}^\ast_1).
 \label{eq:proof4}
\end{equation}
It follows from \eqref{eq:proof3} and \eqref{eq:proof4} that
$J_1(\vc{u}^\ast_1)=J_1(\vc{u}^\ast_0)$,
and hence $\vc{u}^\ast_0$ achieves the minimum value of $J_1$.
That is, $\vc{u}_0^\ast\in\U_1^\ast$ and $\U_0^\ast\subset\U_1^\ast$.
\end{IEEEproof}

Theorem~\ref{thm:L1optimal} suggests that
$L^1$ optimization can be used for 
the maximum hands-off (or the $L^0$-optimal) solution.
The relation between $L^1$ and $L^0$ is analogous to the situation in compressed sensing,
where $\ell^1$ optimality is often used to obtain the sparsest (i.e. $\ell^0$-optimal) vector;
see \cite{Ela,EldKut,HayNagTan13} for details.

Finally, we show that when the system is linear, 
the number of switching in the maximum hands-off control
is bounded.
\begin{prop}
\label{prop:switching}
Suppose that the plant is given by a linear system
\[
 \frac{d\vc{x}(t)}{dt}=A\vc{x}(t)+\sum_{i=1}^m \vc{b}_iu_i(t),
\]
where $A\in{\mathbb{R}}^{n\times n}$ and
$\vc{b}_1,\ldots,\vc{b}_m\in{\mathbb{R}}^n$.
Assume that $(A,\vc{b}_1),\ldots,(A,\vc{b}_m)$ are all controllable and
$A$ is nonsingular.
Assume also that the horizon length $T>0$ (for given initial state $\vc{x}(0)=\vc{\xi}\in{\mathcal{R}}$)
is chosen such that an $L^1$-optimal control exists.
Let $\omega$ be the largest imaginary part of the eigenvalues of $A$.
Then, the maximum hands-off control is a piecewise constant signal,
with values $-1$, $0$, and $1$,
with no switches from $+1$ to $-1$ or $-1$ to $+1$,
and with $2nm(1+T\omega/\pi)$ discontinuities at most.
\end{prop}
\begin{IEEEproof}
Since $(A,\vc{b}_1),\ldots,(A,\vc{b}_m)$ are controllable and
$A$ is nonsingular, the $L^1$-optimal control problem is
normal \cite[Theorem 6-13]{AthFal}.
Then, by Theorem \ref{thm:L1optimal},
the maximum hands-off control is identical to the $L^1$-optimal control.
Combining this with Theorem 3.2 of \cite{Haj79} gives the results.
\end{IEEEproof}

\subsection{Discrete-time hands-off control}
\label{subsec:DT}
Here we consider discrete-time hands-off control.
We assume the plant model is given by
\begin{equation}
 \vc{x}[k+1] = \vc{f}\bigl(\vc{x}[k]\bigr) + \sum_{i=1}^m \vc{g}_i\bigl(\vc{x}[k]\bigr)u_i[k], ~k=0,1,\ldots,N-1,
 \label{eq:DT_plant}
\end{equation}
where
$\vc{x}[k]\in{\mathbb{R}}^n$ is the discrete-time state,
$u_1[k],\dots,u_m[k]$ are the discrete-time scalar control inputs,
$\vc{f}$ and $\vc{g}_i$
are functions on $\R^n$.
We assume that $\vc{f}(\vc{x})$, $\vc{g}_i(\vc{x})$,
$\vc{f}'(\vc{x})$, and $\vc{g}_i'(\vc{x})$
are continuous.
We use the vector notation $\vc{u}[k]\eq[u_1[k],\dots,u_m[k]]^\top$.

The control $\{\vc{u}[0],\vc{u}[1],\ldots,\vc{u}[N-1]\}$ is chosen to drive the state $\vc{x}[k]$
from a given initial state 
$\vc{x}[0]=\vc{\xi}$
to the origin 
$\vc{x}[N]=\vc{0}$.
The components of the control $\vc{u}[k]$ are constrained in magnitude by
\begin{equation}
 \max_i |u_i[k]| \leq 1,~ k=0,1,\ldots,N-1.
 \label{eq:DT_input_constraint}
\end{equation}
We call a control $\{\vc{u}[0],\ldots,\vc{u}[N-1]\}$ admissible
(as in the continuous-time case)
if it satisfies \eqref{eq:DT_input_constraint}
and the resultant state $\vc{x}[k]$ from \eqref{eq:DT_plant} satisfies
$\vc{x}[0]=\vc{\xi}$ and $\vc{x}[N]=\vc{0}$.
We denote by $\U[N,\vc{\xi}]$ the set of all admissible controls.
We assume that $N$ is sufficiently large so that the set $\U[N,\vc{\xi}]$ is non-empty.

For the admissible control, we consider the discrete-time maximum hands-off control
(or $\ell^0$-optimal control) defined by
\begin{equation}
 \underset{\vc{u}\in{\U[N,\vc{\xi}]}}{\mathrm{minimize~}}{J_0(\vc{u})},~
 J_0(\vc{u}) \triangleq \frac{1}{N} \sum_{i=1}^m \lambda_i \|\vc{u}_i\|_{\ell^0},
 \label{eq:DT_l0}
\end{equation}
where $\|\vc{v}\|_{\ell^0}$ denotes the number of the nonzero elements of $\vc{v}\in\mathbb{R}^N$.
The associated $\ell^1$-optimal control problem is given by
\begin{equation}
 \begin{split}
 &\underset{\vc{u}\in{\U[N,\vc{\xi}]}}{\mathrm{minimize~}}{J_1(\vc{u})},\\
 &J_1(\vc{u}) \triangleq \frac{1}{N} \sum_{i=1}^m \lambda_i \|\vc{u}_i\|_{\ell^1}
  =\frac{1}{N} \sum_{i=1}^m \sum_{k=0}^{N-1}\lambda_i |\vc{u}_i[k]|.
 \end{split}
 \label{eq:DT_l1}
\end{equation}
For the $\ell^1$-optimal control problem, we define the Hamiltonian $H(\vc{x},\vc{p},\vc{u})$
by
\[
\begin{split}
 &H(\vc{x},\vc{p},\vc{u}) \triangleq \frac{1}{N}\sum_{i=1}^m \lambda_i |u_i|
 + \vc{p}^\top \left(\vc{f}\bigl(\vc{x}\bigr)+\sum_{i=1}^m\vc{g}_i\bigl(\vc{x}\bigr)u_i\right),
 \end{split}	
\]
where $\vc{p}$ denotes the costate for the $\ell^1$-optimal control problem.
Let $\vc{u}^\ast$ be an $\ell^1$-optimal control, and
$\vc{x}^\ast$ and $\vc{p}^\ast$ are the associated state and costate, respectively.
Then the discrete-time minimum principle \cite{Fri} gives
\[
 H(\vc{x}^\ast[k],\vc{p}^\ast[k+1],\vc{u}^\ast[k]) \leq H(\vc{x}^\ast[k],\vc{p}^\ast[k+1],\vc{u}[k]),
\]
for $k=0,1,\ldots,N-1$ and all admissible $\vc{u}\in\U[N,\vc{\xi}]$.
From this, the $\ell^1$-optimal control $\vc{u}_i^\ast$ (if it exists) satisfies
\[
 \vc{u}_i^\ast[k] = -D_{\lambda_i/N}\biggl(\vc{g}_i\bigl(\vc{x}^\ast[k]\bigr)^\top \vc{p}^\ast[k+1]\biggr),
\]
where $D_{\lambda}(\cdot)$ is the dead-zone function defined in \eqref{eq:dez} (see also Fig.~\ref{fig:dez}).
Based on this, we define the discrete-time normality.
\begin{defn}[Discrete-time normality]
The discrete-time $\ell^1$-optimal control problem is said to be \emph{normal}
if
\[
 \bigl|N\lambda_i^{-1}\vc{g}_i\bigl(\vc{x}^\ast[k]\bigr)^\top\vc{p}^\ast[k+1]\bigr|\neq 1,
\]
for $k=0,1,\ldots,N-1$.
\end{defn}
Then we have the following result:
\begin{thm}
\label{thm:DT_l1optimal}
Assume that the discrete-time $\ell^1$-optimal control problem
described in \eqref{eq:DT_l1} is normal
and has at least one solution.
Let $\U_0^\ast$ and $\U_1^\ast$ be the sets of the solutions
of the maximum hands-off control problem in \eqref{eq:DT_l0}
and the $\ell^1$-optimal control problem in \eqref{eq:DT_l1},
respectively.
Then we have $\U_0^\ast=\U_1^\ast$.
\end{thm}
\begin{IEEEproof}
The theorem can be proved using the same ideas used in
the proof of Theorem \ref{thm:L1optimal}.
Details are omitted for sake of brevity.
\end{IEEEproof}

\section{$L^1$/$L^2$-Optimal Control}
\label{sec:L1L2}
In the previous section, we have shown that
the maximum hands-off control problem can be solved
via $L^1$-optimal control.
From the "bang-off-bang" property of the $L^1$-optimal control,
the control changes its value at switching times \emph{discontinuously}.
This is undesirable for some applications in which the actuators cannot move
abruptly.
In this case, one may want to make the control \emph{continuous}.
For this purpose, we add a regularization term to the $L^1$ cost $J_1(\vc{u})$
defined in \eqref{eq:J_L1}.
More precisely, we consider the following mixed $L^1$/$L^2$-optimal control problem.
\begin{problem}[$L^1$/$L^2$-optimal control]
\label{prob:L1L2}
Find an admissible control on $[0,T]$, $\vc{u}\in\U(T,\vc{\xi})$,  that minimizes
\begin{equation}
 \begin{split}
  J_{12}(\vc{u}) 
  &\eq 
   \frac{1}{T}\sum_{i=1}^m \biggl(\lambda_i\|u_i\|_{1}+\frac{\theta_i}{2}\|u_i\|_{2}^2\biggr)\\
  &=  
   \frac{1}{T}\sum_{i=1}^m \int_0^T \biggl(\lambda_i |u_i(t)| + \frac{\theta_i}{2}|u_i(t)|^2\biggr) dt,
 \end{split}
 \label{eq:J_12}
\end{equation}
where $\lambda_i>0$ and $\theta_i>0$, $i=1,\dots,m$, are given weights.
\hfill\IEEEQEDhere
\end{problem}

To discuss the optimal solution(s) of the above problem,
we next give necessary conditions for the $L^1$/$L^2$-optimal control
using the minimum principle of Pontryagin.

The Hamiltonian function associated to Problem~\ref{prob:L1L2} is given by
\[
 \begin{split}
  H(\vc{x},\vc{p},\vc{u})
   &= \sum_{i=1}^m \biggl(\lambda_i|u_i|+\frac{\theta_i}{2}|u_i|^2\biggr)\\
    &\quad + \vc{p}^\top \biggl(\vc{f}(\vc{x})+\sum_{i=1}^m\vc{g}_i(\vc{x})u_i\biggr)
 \end{split}
\]
where $\vc{p}$ is the costate vector.
Let $\vc{u}^\ast$ denote the optimal control and
$\vc{x}^\ast$ and $\vc{p}^\ast$ the resultant optimal state and costate,
respectively.
Then we have the following result.
\begin{lem}
\label{lem:L1L2}
The $i$-th element $u_i^\ast(t)$ of
the $L^1$/$L^2$-optimal control $\vc{u}^\ast(t)$
satisfies
\begin{equation}
 u_i^\ast(t) = -\sat\left\{\shrink_{\lambda_i/\theta_i}\left(\theta_i^{-1}\vc{g}_i\bigl(\vc{x}^\ast(t)\bigr)^\top\vc{p}^\ast(t)\right)\right\},
 \label{eq:uopt_1}
\end{equation}
where 
$\shrink_{\lambda/\theta}(\cdot)$ is the shrinkage function defined by
\[
 \shrink_{\lambda/\theta}(v)  \eq
   \begin{cases} 
     v+\lambda/\theta& {\rm{if~}} v<-\lambda/\theta,\\
     0, & {\rm{if~}} -\lambda/\theta\leq v \leq \lambda/\theta,\\
     v-\lambda/\theta, & {\rm{if~}} \lambda/\theta<v,\\
   \end{cases}
\]
and
$\sat(\cdot)$ is the saturation function defined by
\[
 \sat(v)  \eq
   \begin{cases} 
     -1, & \text{\rm if~} v<-1,\\
     v, & \text{\rm if~} -1\leq v \leq 1,\\
     1, & \text{\rm if~} 1<v.\\
   \end{cases}
\]
See Figs.~\ref{fig:shrink} and \ref{fig:sat_shrink} for the graphs of
$\shrink_{\lambda/\theta}(\cdot)$ and 
$\sat\!\left(\shrink_{\lambda/\theta}(\cdot)\right)$,
respectively.
\begin{figure}[tb]
\centering
\includegraphics[width=\linewidth]{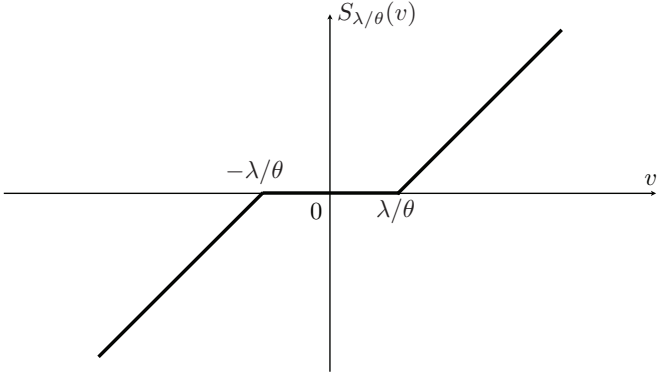}
\caption{Shrinkage function $\shrink_{\lambda/\theta}(v)$}
\label{fig:shrink}
\end{figure}
\begin{figure}[tb]
\centering
\includegraphics[width=\linewidth]{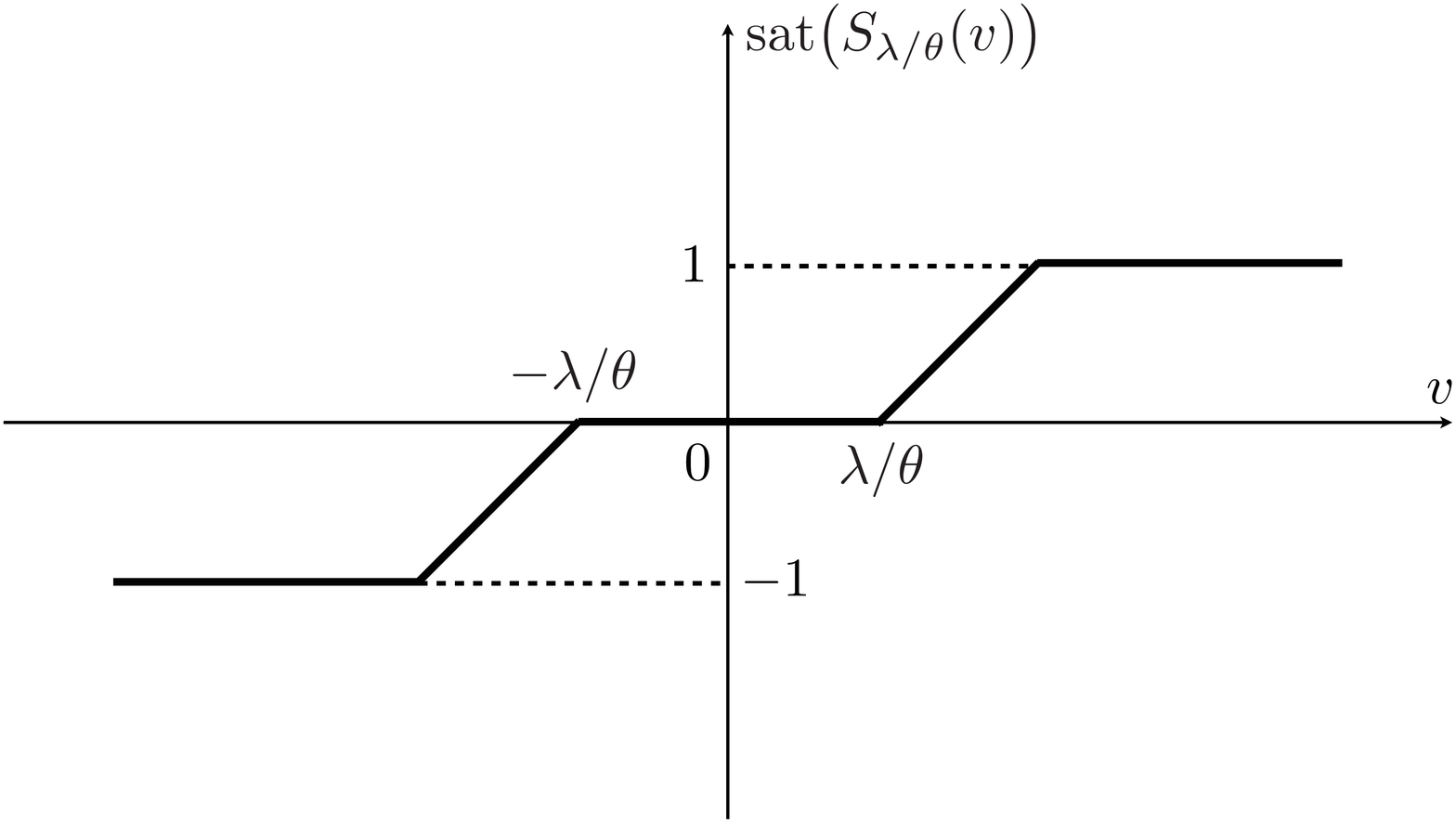}
\caption{Saturated shrinkage function $\sat\!\left(\shrink_{\lambda/\theta}(v)\right)$}
\label{fig:sat_shrink}
\end{figure}
\end{lem}
\begin{IEEEproof}
The result is easily obtained upon noting that
\[
 -\sat\left\{\shrink_{\lambda/\theta}\left(\theta^{-1}a\right)\right\} = \argmin_{|u|\leq 1} \lambda |u| 
  + \frac{\theta}{2}|u|^2+au,
\]
for any $\lambda>0$, $\theta>0$, and $a\in\R$.
\end{IEEEproof}

From Lemma~\ref{lem:L1L2}, we have the following proposition.
\begin{prop}[Continuity]
\label{prop:continuity}
The $L^1$/$L^2$-optimal control $\vc{u}^\ast(t)$ is continuous in $t$ over $[0,T]$.
\end{prop}
\begin{IEEEproof}
Without loss of generality, we assume $m=1$
(a single input plant),
and omit subscripts for $u$, $\theta$, $\lambda$, and so on.
Let
\[
 \bar{u}(\vc{x},\vc{p}) \eq -\sat\left\{\shrink_{\lambda/\theta}\left(\theta^{-1}\vc{g}(\vc{x})^\top\vc{p}\right)\right\}.
\]
Since functions $\left(\sat\circ\shrink_{\lambda/\theta}\right)(\cdot)$ and $\vc{g}(\cdot)$
are continuous,
$\bar{u}(\vc{x},\vc{p})$ is also continuous in $\vc{x}$ and $\vc{p}$.
It follows from Lemma~\ref{lem:L1L2} that
the optimal control $u^\ast$ given in \eqref{eq:uopt_1}
is continuous in
$\vc{x}^\ast$ and $\vc{p}^\ast$.
Hence, $u^\ast(t)$ is continuous, if $\vc{x}^\ast(t)$ and $\vc{p}^\ast(t)$
are continuous in $t$ over $[0,T]$.

The canonical system for the $L^1$/$L^2$-optimal control is given by
\[
 \begin{split}
  \frac{d\vc{x}^\ast(t)}{dt} &=  \vc{f}\bigl(\vc{x}^\ast(t)\bigr) + \vc{g}\bigl(\vc{x}^\ast(t)\bigr)\bar{u}\bigl(\vc{x}^\ast(t),\vc{p}^\ast(t)\bigr),\\
  \frac{d\vc{p}^\ast(t)}{dt} &= - \vc{f}'\bigl(\vc{x}^\ast(t)\bigr)^\top \vc{p}^\ast(t)\\
  	&\qquad - \bar{u}\bigl(\vc{x}^\ast(t),\vc{p}^\ast(t)\bigr) \vc{g}'\bigl(\vc{x}^\ast(t)\bigr)^\top \vc{p}^\ast(t).
 \end{split} 
\]
Since $\vc{f}(\vc{x})$, $\vc{g}(\vc{x})$, $\vc{f}'(\vc{x})$, and $\vc{g}'(\vc{x})$ are continuous
in $\vc{x}$ by assumption,
and so is $\bar{u}(\vc{x},\vc{p})$ in $\vc{x}$ and $\vc{p}$,
the right hand side of the canonical system is continuous in $\vc{x}^\ast$ and $\vc{p}^\ast$.
From a continuity theorem of dynamical systems,
e.g. \cite[Theorem 3-14]{AthFal},
it follows that the resultant trajectories $\vc{x}^\ast(t)$ and $\vc{p}^\ast(t)$ are
continuous in $t$ over $[0,T]$.
\end{IEEEproof}

Proposition~\ref{prop:continuity} motivates us to use the $L^1/L^2$ optimization
in Problem~\ref{prob:L1L2} for continuous hands-off control.

In general, the degree of continuity (or smoothness) and the sparsity of the control input
cannot be optimized at the same time.
The weights $\lambda_i$ or $\theta_i$ can be used for
trading smoothness for sparsity.
Lemma~\ref{lem:L1L2} suggests that
increasing the weight $\lambda_i$ (or decreasing $\theta_i$)
makes the $i$-th input $u_i(t)$ sparser
(see also Fig.~\ref{fig:sat_shrink}).
On the other hand,
decreasing $\lambda_i$ (or increasing $\theta_i$)
smoothens $u_i(t)$.
In fact, we have the following limiting properties.
\begin{prop}[Limiting cases]
Assume the $L^1$-optimal control problem 
is normal.
Let $\vc{u}_1(\vc{\lambda})$ and $\vc{u}_{12}(\vc{\lambda},\vc{\theta})$
be solutions to respectively Problems~\ref{prob:L1} and \ref{prob:L1L2}
with parameters
\[
 \vc{\lambda}\eq(\lambda_1,\dots,\lambda_m),\quad
 \vc{\theta}\eq(\theta_1,\dots,\theta_m).
\]
\begin{enumerate}
\item For any fixed $\vc{\lambda}>0$,
we have
\[
 \lim_{\vc{\theta}\to\vc{0}} \vc{u}_{12}(\vc{\lambda},\vc{\theta}) = \vc{u}_1(\vc{\lambda}).
\] 
\item For any fixed $\vc{\theta}>0$,
we have
\[
 \lim_{\vc{\lambda}\to\vc{0}}\vc{u}_{12}(\vc{\lambda},\vc{\theta}) = \vc{u}_2(\vc{\theta}),
\]
where $\vc{u}_2(\vc{\theta})$ is an $L^2$-optimal (or minimum energy) control
discussed in \cite[Chap.~6]{AthFal},
that is,
a solution to a control problem where $J_1(\vc{u})$ in Problem~\ref{prob:L1} is replaced with
\begin{equation}
 J_2(\vc{u}) = \frac{1}{T} \sum_{i=1}^m \frac{\theta_i}{2}\int_0^T|u_i(t)|^2dt.
 \label{eq:J2}
\end{equation}
\end{enumerate}
\end{prop}
\begin{IEEEproof}
The first statement follows directly from the fact that
for any fixed $\lambda>0$, we have
\[
 \lim_{\theta\to 0}\sat\!\left({\shrink}_{\lambda/\theta}(\theta^{-1}w)\right) = \dez_{\lambda}(w),\quad
 \forall w\in\R\setminus\{\pm \lambda\},
\]
where $\dez_\lambda(\cdot)$ is the dead-zone function defined in \eqref{eq:dez}. 
The second statement derives from the fact that
for any fixed $\theta>0$, we have
\[
 \lim_{\lambda\to 0}\sat\!\left(\shrink_{\lambda/\theta}(v)\right) = \sat(v),\quad
 \forall v\in\R.
\]
\end{IEEEproof}

In summary, the $L^1$/$L^2$-optimal control is an \emph{intermediate control} between 
the $L^1$-optimal control (or the maximum hands-off control)
and the $L^2$-optimal control.

\begin{example}
Let us consider the following linear system
\[
 \frac{d\vc{x}(t)}{dt} =
  \begin{bmatrix}0&-1&0&0\\1&0&0&0\\0&1&0&0\\0&0&1&0\end{bmatrix}\vc{x}(t)
  + \begin{bmatrix}2\\0\\0\\0\end{bmatrix}u(t).
\]
We set the final time $T=10$, and the initial and final states as
\[
 \vc{x}(0) = [1,1,1,1]^\top,\quad \vc{x}(10)=\vc{0}.
\] 
Fig.~\ref{fig:L1L2} 
shows the $L^1$/$L^2$ optimal control with
weights $\lambda_1=\theta_1=1$.
The maximum hands-off control is also illustrated.
\begin{figure}[tbp]
\centering
\includegraphics[width=\linewidth]{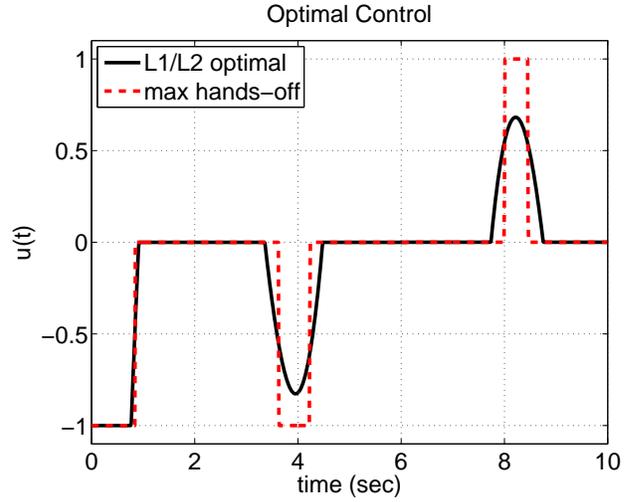}
\caption{Maximum hands-off control (dashed) and  $L^1$/$L^2$-optimal control (solid)}
\label{fig:L1L2}
\end{figure}
We can see that the $L^1$/$L^2$-optimal control is continuous but sufficiently sparse.
Fig.~\ref{fig:L1L2state} shows the state trajectories of $x_i(t)$, $i=1,2,3,4$.
By the sparse $L^1$/$L^2$ control, each state approaches zero within time $T=10$.
\begin{figure}[tbp]
\centering
\includegraphics[width=\linewidth]{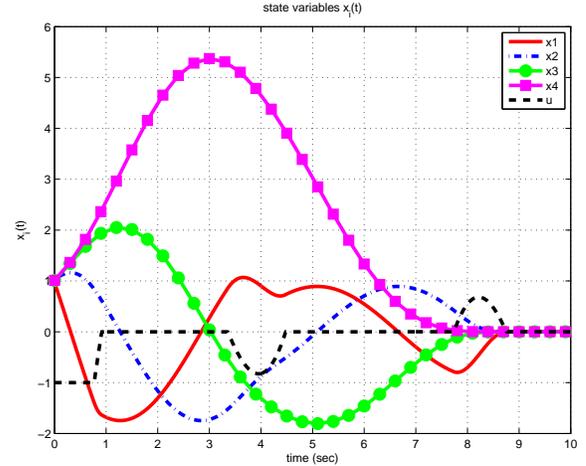}
\caption{State trajectory by $L^1$/$L^2$-optimal control}
\label{fig:L1L2state}
\end{figure}
\end{example}

\section{Self-Triggered Hands-Off Feedback Control}
\label{sec:feedback}
In the previous section, we have shown that the maximum hands-off control
is given by the solution to an associated $L^1$-optimal control problem.
The $L^1$-optimal control can be computed, for example, via convex optimization after
time discretization.
However, it is still difficult to give optimal control as a function of the state variable $\vc{x}(t)$.
This is a drawback if there exist uncertainties in the plant model and disturbances added to the signals.
Therefore, we extend maximum hands-off control to feedback control.
In this section, we assume the controlled plant model is given by
a single-input, linear time-invariant system
\begin{equation}
 \frac{d\vc{x}(t)}{dt} = A \vc{x}(t) + \vc{b}u(t) +\vc{d}(t),\quad t\in [0,\infty),
 \label{eq:plant-linear}
\end{equation}
where $A\in\R^{n\times n}$ and $\vc{b}\in\R^{n}$
are given constants, and $\vc{d}(t)\in\R^n$ denotes 
an unknown plant disturbance.
For a nonlinear plant, one can use \eqref{eq:plant-linear}
as a linearized model and $d(t)$ as the linearization error
(see Section \ref{sec:examples}).
We assume that
\begin{enumerate}
 \item $(A,\vc{b})$ is reachable,
 \item $A$ is nonsingular.
\end{enumerate}
This is a sufficient condition so that the $L^1$-optimal control problem with
the single-input linear system
\eqref{eq:plant-linear} in the disturbance-free case where $\vc{d}\equiv\vc{0}$ is
normal for any horizon length $T>0$ and any initial condition $\vc{x}(0)\in{\mathcal R}$
\cite[Theorem 6-13]{AthFal}.

\subsection{Sparsity Rate for Infinite Horizon Signals}
\label{subsec:sparsity-rate}
Before considering feedback control, we define the sparsity rate
for infinite horizon signals
(cf. Definition~\ref{defn:sparsity-rate-f}).
\begin{defn}[Sparsity rate]
For infinite horizon signal $u=\{u(t):t\in[0,\infty)\}$, we define
the sparsity rate by
\begin{equation}
 \spr_\infty(u) \eq \lim_{T\to\infty} \frac{1}{T} \left\|u|_{[0,T]}\right\|_{0},
\end{equation}
\end{defn}
where $u|_{[0,T]}$ is the restriction of $u$ to the interval $[0,T]$.
Note that
\begin{enumerate}
\item If $\|u\|_{0}<\infty$,
then $\spr_\infty(u)=0$.
\item If $|u(t)|>0$ for almost all $t\in [0,\infty)$, then $\spr_\infty(u)=1$.
\item For any measurable function $u$ on $[0,\infty)$, we have
$0\leq \spr_\infty(u) \leq 1$.
\end{enumerate}
We say again that an infinite horizon signal $u$ is \emph{sparse}
if the sparsity rate $\spr_\infty(u)\ll 1$.

\begin{lem}
\label{lem:sparsity-rate}
Let $u$ be a measurable function on $[0,\infty)$.
If there exist time instants $t_0, t_1,t_2,\dots$
such that
\[
 \begin{split}
  &t_0=0,~ t_{k+1}=t_k+T_k,~
  T_k>0,\\
  &\spr_{T_k}\bigl(u|_{[t_k,t_{k+1}]}\bigr) \leq r, \quad \forall k \in \{0,1,2,\dots\},
 \end{split}
\] 
then $\spr_\infty(u) \leq r$.
\end{lem}
\begin{IEEEproof}
The following calculation proves the statement.
\[
 \begin{split}
  \spr_\infty(u) &= \lim_{T\to\infty}\frac{1}{T} \left\| u|_{[0,T]} \right\|_{0}\\
   &= \lim_{N\to\infty}\frac{1}{t_N}\sum_{k=0}^{N-1}\left\|u|_{[t_k,t_{k+1}]}\right\|_{0}\\
   &= \lim_{N\to\infty}\frac{1}{t_N}\sum_{k=0}^{N-1} (t_{k+1}-t_k) \spr_{T_k}\bigl(u|_{[t_k,t_{k+1}]}\bigr)\\
   &\leq \lim_{N\to\infty}\frac{1}{t_N} (t_N-t_0) r\\
   &= r
  \end{split}
\]   
\end{IEEEproof}

\subsection{Control Algorithm}
\label{subsec:control-algorithm}
Fix a bound on the sparsity rate $\spr_\infty(u)\leq r$ with $r\in (0,1)$.
We here propose a feedback control algorithm
that achieves the sparsity rate $r$ of the resultant control input.
Our method involves applying maximum hands-off control over finite horizons,
and to use self-triggered feedback to compensate for disturbances.
In self-triggered control, the next update time is determined by
the current plant state.

First, let us assume that an initial state $\vc{x}(0)=\vc{x}_0\in \R^n$ is given.
For this, we compute the minimum-time $T^\ast(\vc{x}_0)$, the solution
of the minimum-time control.
Then, we define the first sampling period (or the first horizon length) by
\[
 T_0 \eq \max\left\{T_{\min},r^{-1}T^\ast(\vc{x}_0)\right\},
\]
where $T_{\min}$ is a given positive time length that prevents
the sampling period from zero
(thereby avoiding 
Zeno executions \cite{ZhaJohLygSas01}).
For this horizon length, we compute the maximum hands-off control
on the interval $[0,T_0]$.
Let this optimal control be denoted $u_0(t)$, $t\in[0,T_0]$, that is
\[
 u_0(t) = \argmin_{u\in \U(T_0,\vc{x}_0)} \|u\|_{0},\quad t \in [0,T_0],
\]
where $\U(T_0,\vc{x}_0)$ is the set of admissible control on
time interval $[0,T_0]$ with initial state $\vc{x}_0$;
see Section~\ref{sec:MHOC-problem}.
Apply this control, $u_0(t)$, to the plant
\eqref{eq:plant-linear}
from $t=0$ to $t=T_0$.
If $d\equiv 0$ (i.e. no disturbances), then $\vc{x}(T_0)=0$
by the terminal constraint, and applying $u(t)=0$ for $t\geq T_0$
gives
$\vc{x}(t)=0$ for all $t\geq T_0$.

However, if $d\not\equiv 0$, then $\vc{x}(T_0)$ will in general not be exactly zero.
To steer the state to the origin, we should again apply a control to the plant.
Let $\vc{x}_1 \eq \vc{x}(T_0)$,
and $t_1\eq T_0$.
We propose to compute the minimum time $T^\ast(\vc{x}_1)$ and let
\[
 T_1 \eq \max\left\{T_{\min},r^{-1}T^\ast(\vc{x}_1)\right\}.
\]
For this horizon length $T_1$, we compute the maximum hands-off control,
$u_1(t)$, $t\in[t_1, t_1+T_1]$, as well,
which is applied to the plant on the time interval $[t_1, t_1+T_1]$.

Continuing this process gives a self-triggered feedback control algorithm,
described in Algorithm~\ref{alg:feedback},
which results in an infinite horizon control
\begin{equation}
 u(t) = u_k(t-t_k),\quad t\in [t_k,t_{k+1}],\quad k=0,1,2,\dots,
 \label{eq:infcontrol}
\end{equation}
where $u_k$ is defined in \eqref{eq:uk}.
\begin{algorithm}[tb]
\caption{Self-triggered Hands-off Control}
\label{alg:feedback}
\begin{algorithmic}
 \State Given initial state $\vc{x}_0$ and minimum inter-sampling time $T_{\min}$.
 \State Let $\vc{x}(0)=\vc{x}_0$ and $t_0=0$.
 \For {$k=0,1,2,\dots$}
 	\State Measure $\vc{x}_k := \vc{x}(t_k)$.
	\State Compute $T^\ast(\vc{x}_k)$.
	\State Put $T_k:=\max\left\{T_{\min}, r^{-1}T^\ast(\vc{x}_k)\right\}$.
	\State Put $t_{k+1} := t_k + T_k$.
	\State Compute max hands-off control
		\begin{equation}
		  u_k = \argmin_{u\in\U(T_k,\vc{x}_k)} \|u\|_{0}.
		  \label{eq:uk}
		\end{equation}  
	\State Apply $u(t)=u_k(t-t_k),~ t\in[t_k,t_{k+1}]$ to the plant.
 \EndFor
\end{algorithmic}
\end{algorithm}
For this control, we have the following proposition.
\begin{prop}[Sparsity rate]
 For the infinite horizon control $u$ in \eqref{eq:infcontrol},
 the sparsity rate $\spr_\infty(u)$ is less than $r$.
\end{prop}
\begin{IEEEproof}
Fix $k\in\{0,1,2,\dots\}$.
Let $\vc{x}_k \eq \vc{x}(t_k)$.
The $k$-th horizon length $T_k$ is given by
\begin{equation}
 T_k = \max\left\{T_{\min}, r^{-1}T^\ast(\vc{x}_k)\right\}.
 \label{eq:Tk}
\end{equation}
Let us first consider the case when $T_{\min}\leq r^{-1}T^\ast(\vc{x}_k)$, or $T_k=r^{-1}T^\ast(\vc{x}_k)$.
Let $u^\ast_k(t)$ denote the minimum-time control for initial state $\vc{x}_k$,
and define
\begin{equation}
 \tilde{u}_k(t) := \begin{cases}u^\ast_k(t), & t\in[0,T^\ast(\vc{x}_k)],\\
 	0, & t \in (T^\ast(\vc{x}_k),r^{-1}T^\ast(\vc{x})].
	\end{cases}
 \label{eq:tuk}	
\end{equation}
Note that $T^\ast(\vc{x}_k)<r^{-1}T^\ast(\vc{x}_k)$ since $r\in(0,1)$.
Clearly this is an admissible control, that is, 
$\tilde{u}_k \in \U(T_k, \vc{x}_k)$,
and
\[
 \|\tilde{u}_k\|_{0} = \|u^\ast_k\|_{0} = T^\ast(\vc{x}_k),
\]
for which see also Fig.~\ref{fig:admissible_control}.
\begin{figure}[tb]
\centering
\includegraphics[width=\linewidth]{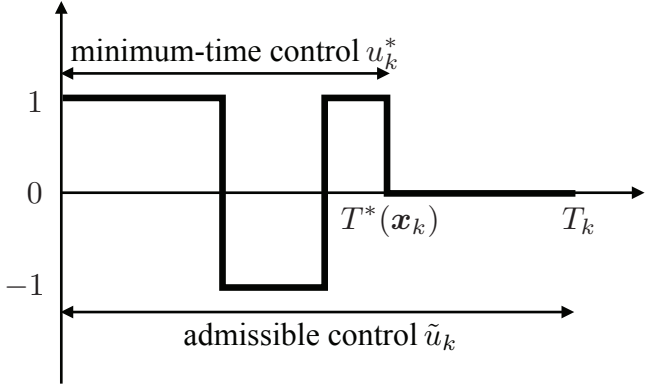}
\caption{
Minimum-time control $u^\ast_k(t)$ and admissible control $\tilde{u}_k(t)$
defined in \eqref{eq:tuk}.
}
\label{fig:admissible_control}
\end{figure}
On the other hand, let $u_k$ denote the maximum hands-off control
on time interval $[0,T_k]$ with initial state $\vc{x}_k$.
Since $u_k$ has the minimum $L^0$ norm, we have
\[
 \|u_k\|_{0} \leq \|\tilde{u}_k\|_{0} = T^\ast(\vc{x}_k).
\] 
It follows that the sparsity rate of $u_k(t-t_k)$, $t\in[t_k,t_k+T_k]$ is
\[
 \spr_{T_k}(u_k) = \frac{1}{T_k}\|u_k\|_{0} \leq \frac{T^\ast(\vc{x}_k)}{r^{-1}T^\ast(\vc{x}_k)} = r.
\]
Next, for the case when $T_{\min}\geq r^{-1}T^\ast(\vc{x}_k)$,
we have $T_{\min} > T^\ast(\vc{x})$.
It follows that $\spr_{T_k}(u_k) \leq r$ by a similar argument.
In either case, we have $\spr_{T_k}(u_k)\leq r$ for $k=0,1,2,\dots$.
Finally, Lemma~\ref{lem:sparsity-rate} gives the result.
\end{IEEEproof}

\begin{rem}[Minimum time computation]
Algorithm~\ref{alg:feedback} includes computation of
the minimum time $T^\ast(\vc{x}_k)$.
For single-input, linear time-invariant system,
an efficient numerical algorithm has been proposed in
\cite{ChuWu92}, which one can use for the computation.
Also, this can be used to check whether the initial state $\vc{x}_0$ lies in
the reachable set ${\mathcal R}$.
\end{rem}

\subsection{Practical Stability}
\label{subsec:stability}
By the feedback control algorithm (Algorithm~\ref{alg:feedback}),
the state $\vc{x}(t)$ is sampled at sampling instants $t_k$, $k=1,2,\dots$,
and between sampling instants the system acts as an open loop system.
Since there exists disturbance $\vc{d}(t)$, it is impossible to asymptotically stabilize
the feedback system to the origin.
We thus focus on \emph{practical stability} of the feedback control system
under bounded disturbances.
The following are fundamental lemmas to prove the stability.
\begin{lem}
For $A\in \R^{n\times n}$,
we have
\[
 \bigl\|e^{At}\bigr\| \leq e^{\mu(A) t},\quad \forall t\in [0,\infty),
\]
where $\mu(A)$ is the maximum eigenvalue of $(A+A^\top)/2$, that is,
\begin{equation}
 \mu(A) = \lambda_{\max}\left(\frac{A+A^\top}{2}\right).
 \label{eq:matrix-measure}
\end{equation}
\end{lem}
\begin{IEEEproof}
This can be easily proved by a general theorem of
the matrix measure \cite[Theorem II.8.27]{DesVid}.
\end{IEEEproof}
\begin{lem}
\label{lem:alpha}
There exists a scalar-valued, continuous, and non-decreasing function
$\alpha:[0,\infty) \rightarrow [0,\infty)$ such that
\begin{enumerate}
 \item $\alpha(0)=0$,
 \item $T^\ast(\vc{x})\leq \alpha(\|\vc{x}\|)$, $\forall \vc{x}\in {\mathcal R}$,
 where ${\mathcal R}$ is the reachable set defined in Definition~\ref{defn:reachable-set}
\end{enumerate} 
\end{lem}
\begin{IEEEproof}
For $v\geq 0$, define
\[
 \alpha(v) \eq \max_{\|\vc{\xi}\|\leq v} T^\ast(\vc{\xi}).
\]
By this definition, it is easy to see that if $v_1\geq v_2$ then $\alpha(v_1)\geq \alpha(v_2)$.
Since $T^\ast(\vc{\xi})$ is continuous on ${\mathcal R}$
(see \cite{Haj71}), $\alpha(v)$ is continuous.
The first statement is a result from $T^\ast(\vc{0})=0$.
Then, setting $v=\|\vc{x}\|$ for $\vc{x}\in{\mathcal R}$ gives the second statement.
\end{IEEEproof}

Now, we have the following stability theorem.
\begin{thm}
\label{thm:stability}
Assume that the plant noise is bounded by $\delta>0$, that is,
$\|\vc{d}(t)\| \leq \delta$ for all $t\geq 0$.
Assume also that the initial state $\vc{x}(0)=\vc{x}_0$ is in the reachable set ${\mathcal R}$,
and let
\begin{equation}
 T_0 \eq \max\{T_{\min},r^{-1}T^\ast(\vc{x}_0)\}.
 \label{eq:T0}
\end{equation}
Define
\begin{equation}
 \begin{split}
  \Omega &\eq \left\{\vc{x}\in\R^n: \|\vc{x}\|\leq \gamma \right\},\\
  \gamma &\eq \frac{\delta}{\mu(A)}\left(e^{\mu(A) T_0}-1\right),
 \end{split}
 \label{eq:Omega}
\end{equation}
and assume $\Omega\subset{\mathcal R}$.
Choose a function $\alpha$ which satisfies the conditions in Lemma~\ref{lem:alpha}.
If
\begin{equation}
 \alpha(\gamma) \leq rT_0, \label{eq:alpha-gamma}
\end{equation}
then the feedback control with Algorithm~\ref{alg:feedback} achieves practical stability in the sense that
\begin{enumerate}
\item $\vc{x}(t)$ is bounded for $t\in[0,t_1]$.
\item $\vc{x}_k \eq \vc{x}(t_k)\in \Omega$, $\forall k\in\{1,2,\dots\}$.
\item For $t\in [t_k,t_{k+1}]$, $k\in\{1,2,\dots\}$,
we have $\|\vc{x}(t)\|\leq h$, where if $\mu(A)<0$
\[
 h =  \gamma + \frac{\|\vc{b}\|+\delta}{|\mu(A)|} \eq h_1,
\]
and if $\mu(A)>0$
\[
 h = h_1 e^{\mu(A)\max\{T_{\min},r^{-1}\alpha(\gamma)\}} - \frac{\|\vc{b}\|+\delta}{\mu(A)}.
\] 
\end{enumerate}
\end{thm}
\begin{IEEEproof}
Since the system is linear time-invariant and $u(t)$ and $\vc{d}(t)$ are bounded,
the state $\vc{x}(t)$ is also bounded on $[0,t_1]$.
For $t=t_1$, we have
\[
 \begin{split}
 \|\vc{x}_1\|=\|\vc{x}(t_1)\| &\leq \int_0^{T_0} \bigl\| e^{A(T_0-\tau)} \bigr\|\delta d\tau\\
   &\leq \int_0^{T_0} e^{\mu(A)(T_0-\tau)}\delta d\tau\\
   &=\frac{\delta}{\mu(A)}\left(e^{\mu(A) T_0}-1\right),
 \end{split} 
\] 
and hence $\vc{x}_1=\vc{x}(t_1)\in \Omega$. 
Note that since $\vc{x}_0\in{\mathcal R}$, 
we have $T_0<\infty$. Note also that since $A$ is nonsingular, $\mu(A)\neq 0$.
Fix $k\in \{1,2,\dots\}$, and assume $\vc{x}_k=\vc{x}(t_k)\in \Omega$.
Then we have
\[
 \|\vc{x}_{k+1}\| \leq \frac{\delta}{\mu(A)}\left(e^{\mu(A) T_k}-1\right),
\]
where $T_k$ is as in \eqref{eq:Tk}.
Note that $T_k<\infty$ since $\vc{x}_k\in\Omega\subset{\mathcal{R}}$.
If $T_k = T_{\min}$ then
\[
 \begin{split}
  \|\vc{x}_{k+1}\| &\leq \frac{\delta}{\mu(A)}\left(e^{\mu(A) T_{\min}}-1\right)\\
   &\leq \frac{\delta}{\mu(A)}\left(e^{\mu(A) T_0}-1\right)=\gamma
 \end{split} 
\]
since $T_0 \geq T_{\min}$. 
On the other hand, if $T_k = r^{-1}T^\ast(\vc{x}_k)$ then
\begin{equation}
 \begin{split}
 \|\vc{x}_{k+1}\| &\leq \frac{\delta}{\mu(A)}\left(e^{\mu(A) r^{-1}T^\ast(\vc{x}_k)}-1\right)\\
 &= \gamma +  \frac{\delta}{\mu(A)}\left(e^{\mu(A) r^{-1}T^\ast(\vc{x}_k)}-e^{\mu(A) T_0}\right).
 \end{split}
 \label{eq:xtk-bound}
\end{equation}
Lemma~\ref{lem:alpha}, assumption  $\vc{x}_k\in\Omega$,
and equation \eqref{eq:alpha-gamma} give
\begin{equation}
 T^\ast(\vc{x}_k) \leq \alpha(\|\vc{x}_k\|) \leq \alpha(\gamma) \leq rT_0,
 \label{eq:Tast-bound}
\end{equation}
and hence
\[
 e^{\mu(A) r^{-1}T^\ast(\vc{x}_k)}-e^{\mu(A) T_0} \leq 0.
\]
From \eqref{eq:xtk-bound}, we have $\|\vc{x}_{k+1}\|\leq \gamma$.
In each case, we have $\vc{x}_{k+1}=\vc{x}(t_{k+1})\in \Omega$.

Then, let us consider the intersample behavior of $\vc{x}(t)$, $t\in[t_k,t_{k+1}]$
for $k=1,2,\dots$.
As proved above, we have $\vc{x}_k=\vc{x}(t_k)\in \Omega$.
This gives
\[
 \begin{split}
  \|\vc{x}(t)\| & \leq \bigl\|e^{A(t-t_k)}\bigr\| \|\vc{x}_k\|
   + \int_{t_k}^t \bigl\|e^{A(t-\tau)}\bigr\| \|\vc{b}\| |u_k(t)| d\tau\\
   &~ + \int_{t_k}^t \bigl\|e^{A(t-\tau)}\bigr\| \|\vc{d}(\tau)\| d\tau\\
   &\leq e^{\mu(A)(t-t_k)}\|\vc{x}_k\|
    + \int_{t_k}^t e^{\mu(A)(t-\tau)}d\tau (\|\vc{b}\|+\delta)\\
    &= e^{\mu(A)(t-t_k)}\|\vc{x}_k\| + \frac{\|\vc{b}\|+\delta}{\mu(A)}\left(e^{\mu(A)(t-t_k)}-1\right).
  \end{split}
\]
If $\mu(A)<0$ then $\vc{x}(t)$ is bounded as
\[
 \|\vc{x}(t)\|  \leq \|\vc{x}_k\| + \frac{\|\vc{b}\|+\delta}{|\mu(A)|}
  \leq \gamma + \frac{\|\vc{b}\|+\delta}{|\mu(A)|}.
\]
If $\mu(A)>0$ then $\vc{x}(t)$ is again bounded as
\[
 \begin{split}
  \|\vc{x}(t)\|  &\leq e^{\mu(A) T_k} \|\vc{x}_k\| + \frac{\|\vc{b}\|+\delta}{\mu(A)}\left(e^{\mu(A) T_k}-1\right)\\
  &\leq e^{\mu(A)\max\{T_{\min},r^{-1}\alpha(\gamma)\}}\gamma\\
  &~ + \frac{\|\vc{b}\|+\delta}{\mu(A)}\left(e^{\mu(A)\max\{T_{\min},r^{-1}\alpha(\gamma)\}}-1\right).
 \end{split}
\] 
\end{IEEEproof}

From \eqref{eq:T0} and \eqref{eq:Omega}, 
we conclude that the larger the sparsity rate $r$, the smaller the upper bound $\gamma$.
This shows there is a tradeoff between the sparsity rate of control and the performance.
The analysis is deterministic and the bound is for the worst-case disturbance,
but this is reasonably tight in some cases
when a worst-case disturbance is applied to the system,
as shown in the example below.

\section{Example}
\label{sec:examples}
First, let us consider a simple example with
a $1$-dimensional stable plant model
\begin{equation}
 \frac{dx(t)}{dt} = a x(t) + a u(t) + d(t),
 \label{eq:example_plant}
\end{equation}
where $a<0$. 
We assume bounded disturbance, that is, 
there exists $\delta>0$ such that $|d(t)|\leq \delta$ for all $t\geq 0$.
The plant is normal and hence the maximum hands-off control is
given by $L^1$-optimal control thanks to Theorem~\ref{thm:L1optimal}.
In fact, the optimal control $u_k$ in \eqref{eq:uk} is
computed via the minimum principle for $L^1$-optimal control
\cite[Section 6.14]{AthFal} as
\[
 u_k(t) = \begin{cases}0, & t \in [0,\tau),\\ -\sgn(x(t_k)), & t\in[\tau, T_k], \end{cases}
\]
where
\[
 \tau \eq \frac{1}{|a|}\log\left(e^{|a|T_k}-|x(t_k)|\right).
\]
Also, the minimum time function $T^\ast(x)$ is computed as
(see \cite[Example 6-4]{AthFal})
\[
 T^\ast(x) = \frac{1}{|a|} \log(1+|x|),\quad x\in \R.
\]
It follows that the reachable set ${\mathcal R}=\R$,
and the condition $\Omega \subset {\mathcal R}$
in Theorem~\ref{thm:stability}
always holds.
Since 
$A=a\in\R$, we have $\mu(A)=a$
by \eqref{eq:matrix-measure}.
Then, for any $x\in\R$, we have
\[
 T^\ast(x) = \frac{1}{|a|}\log(1+|x|)\leq \frac{|x|}{|a|},
\]
and hence we can choose $\alpha(v)=v/|a|$
for Lemma~\ref{lem:alpha} and Theorem~\ref{thm:stability}.
The stability condition \eqref{eq:alpha-gamma} becomes
\[
 \alpha\left(\frac{\delta}{|a|}(1-e^{a T_0})\right)\leq rT_0
\]
or
$rT_0a^2 \geq \delta (1-e^{a T_0})$. 

\begin{figure}[tbp]
\centering
\includegraphics[width=\linewidth]{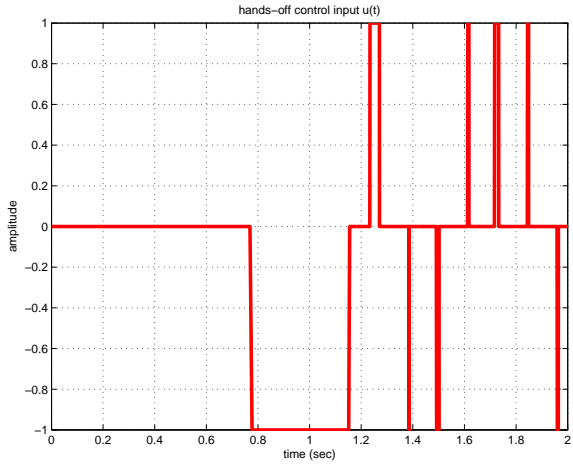}
\caption{Hands-off feedback control with sparsity rate $\spr_\infty(u)=0.148$.}
\label{fig:control}
\end{figure}

For example, with $a=-1$, $\delta=1$, $x_0=1$,
and 
if we choose $T_{\min} < T^\ast(x_0)=\log(1+|x_0|)=\log 2$, then
$rT_0=\log 2$ and the condition becomes
%
\[
 r\geq -\frac{\log 2}{\log(1-\log2)}\approx 0.587.
\]
We set $r=0.6$
and simulate 
the feedback control with disturbance $d(t)$ as 
uniform noise
with mean $0$ and bound $\delta=1$.
Fig.~\ref{fig:control} shows the maximum hands-off control
obtained by Algorithm~\ref{alg:feedback}.
We can observe that the control is sufficiently sparse.
In fact, the sparsity rate for this control is 
$\spr_\infty(u)=0.148$,
which is smaller than the upper bound 
$r=0.6$. 

\begin{figure}[tbp]
\centering
\includegraphics[width=\linewidth]{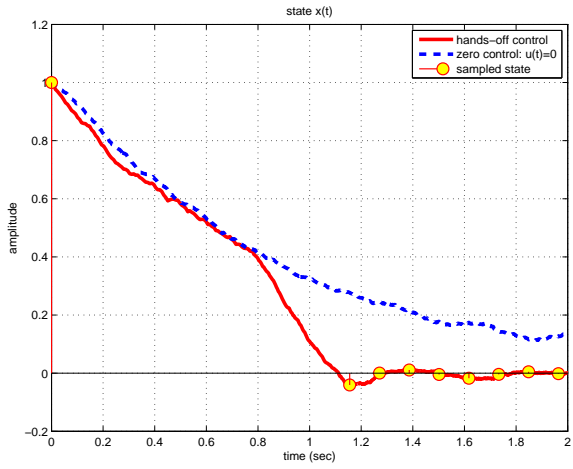}
\caption{
State trajectory: hands-off control (solid) and zero control (dots).
Sampled states $x(t_k)$ are also shown (circles)
}
\label{fig:state}
\end{figure}

Since the plant is asymptotically stable, one can choose
the zero control, that is, $u\equiv 0$, to achieve stability,
which is the sparsest.
Fig.~\ref{fig:state} shows the state $x(t)$ for
the maximum hands-off control and the zero control.
Due to the time optimality of the hands-off control, the state
approaches to 0 faster than that of the zero control.

Then let us consider the influence of disturbances.
The bound $\gamma$ in \eqref{eq:Omega}
is computed as 
$\gamma = 1-\exp(-r^{-1}\log 2)$ 
with $r=0.6$,
and the set $\Omega$ becomes
\[
 \Omega = \left\{x\in \R: |x|\leq 1-\exp(-r^{-1}\log 2)\right\}.
\]
This bound is obtained in a deterministic manner,
and hence the bound is for the worst-case disturbance.
In fact, let us apply a worst-case disturbance
$d(t)=1$ for all $t\geq 0$ to the feedback system.
\begin{figure}[tbp]
\centering
\includegraphics[width=\linewidth]{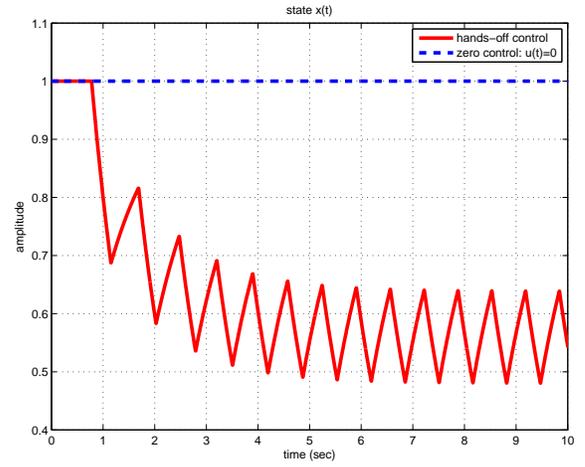}
\caption{State trajectory with worst-case disturbance: hands-off control (solid) and zero control (dots).}
\label{fig:state-worstcase}
\end{figure}
Fig.~\ref{fig:state-worstcase} shows the state trajectories.
The trajectory by the zero control remains $1$ and 
do not approache $0$,
while that by the maximum hands-off control 
still approaches $0$,
and we can see that the bound is reasonably tight.

\begin{figure}[tb]
\centering
\includegraphics[width=\linewidth]{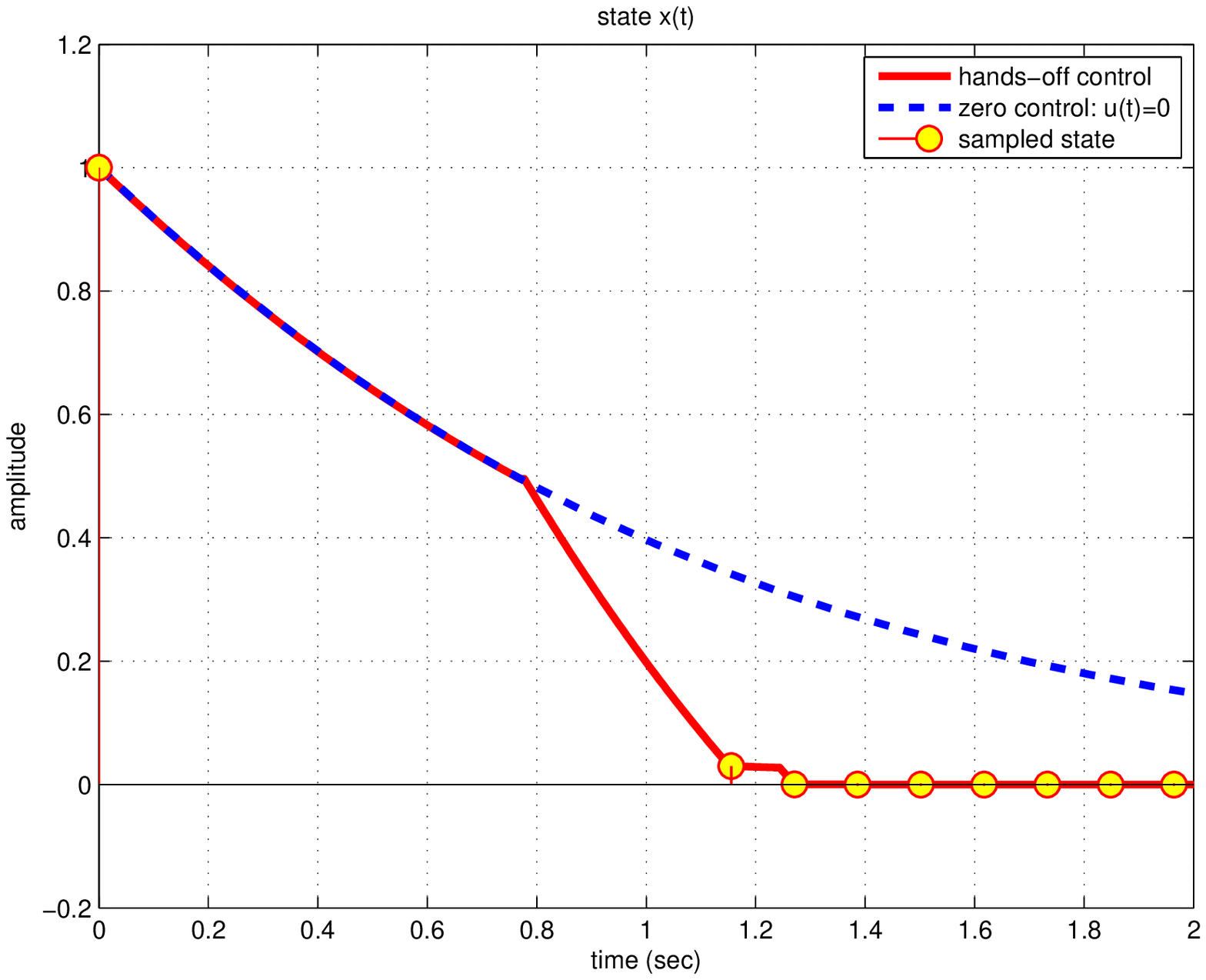}
\caption{State trajectory of nonlinear plant \eqref{eq:example_plant_NL} with $a=-1$ (stable):
ands-off control (solid), zero control (dots),
and sampled states $x(t_k)$ (circles).}
\label{fig:state-NL}
\end{figure}

Next, let us consider a nonlinear plant model
\begin{equation}
 \frac{dx(t)}{dt} = \sin\bigl(ax(t)\bigr) + au(t).
 \label{eq:example_plant_NL}
\end{equation}
We linearize this nonlinear plant to obtain
the linear plant \eqref{eq:example_plant},
with the linearization error
$d(t) \triangleq \sin\big(ax(t)\bigr)-ax(t)$.
Assume $a=-1$ (i.e. stable).
We adopt the control law given as above to the nonlinear plant
\eqref{eq:example_plant_NL}.
Fig.~\ref{fig:state-NL} shows the result.
This figure shows that the hands-off control works well
for the nonlinear plant \eqref{eq:example_plant_NL}.
The sparsity rate of the hands-off control is 
$R_\infty(u)=0.0717$, which is sufficiently small.

\begin{figure}[tb]
\centering
\includegraphics[width=\linewidth]{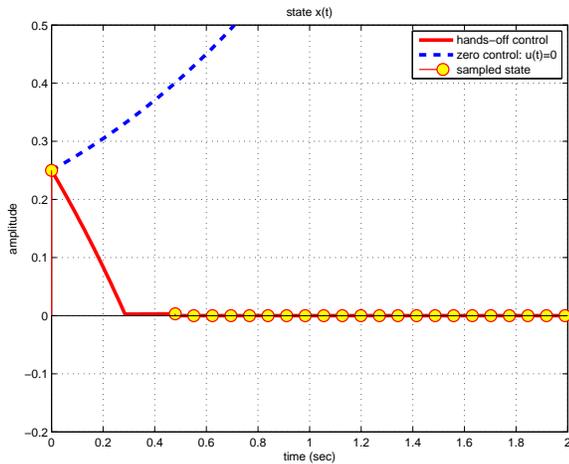}
\caption{
State trajectory of the nonlinear plant \eqref{eq:example_plant_NL} with $a=1$ (unstable): 
hands-off control (solid), zero control (dots),
and sampled states $x(t_k)$ (circles).
}
\label{fig:state-unstable-NL}
\end{figure}

On the other hand, let us consider the nonlinear plant \eqref{eq:example_plant_NL} with
$a=1$ (i.e. unstable). For the linearized plant
\eqref{eq:example_plant_NL}, the hands-off control law is given by
\[
 u_k(t) = \begin{cases}-\sgn(x(t_k)), & t \in [0,\tau),\\ 0, & t\in[\tau, T_k], \end{cases}
\]
where
$\tau \eq -a^{-1}\log\bigl(1-|x(t_k)|\bigr)$.
The minimum time function $T^\ast(x)$ is given by
$T^\ast(x) = -a^{-1} \log\bigl(1-|x|\bigr)$ for $x\in {\mathcal{R}}$,
where ${\mathcal R}=(-1,1)$.
We set the initial state $x_0=0.25$
and the sparsity rate $r=0.6$, and simulate the feedback control
with the nonlinear plant \eqref{eq:example_plant_NL}.
Fig.~\ref{fig:state-unstable-NL} shows
the obtained state trajectory of \eqref{eq:example_plant_NL}.
Obviously, the zero control cannot stabilize the unstable plant
and hence the state diverges, while the hands-off control
keeps the state close to the origin.
The sparsity rate is $R_\infty(u)=0.1135$, which is sufficiently small.

\section{Conclusion}
\label{sec:conclusion}

In this paper, 
we have proposed maximum hands-off control.
It has the minimum support per unit time,
or is the sparsest,
among all admissible controls.
Under normality assumptions, the maximum hands-off control can be computed via
$L^1$-optimal control.
For linear systems, we have also proposed a feedback control algorithm,
which guarantees a given sparsity rate and practical stability.
An example has illustrated the effectiveness of the proposed control.
Future work includes the development of an effective computation algorithm for maximum hands-off control,
for situations when the control problem does not satisfy normality conditions,
and also when the plant is nonlinear.


\begin{IEEEbiography}[
{\includegraphics[width=1in,height=1.25in,clip,keepaspectratio]{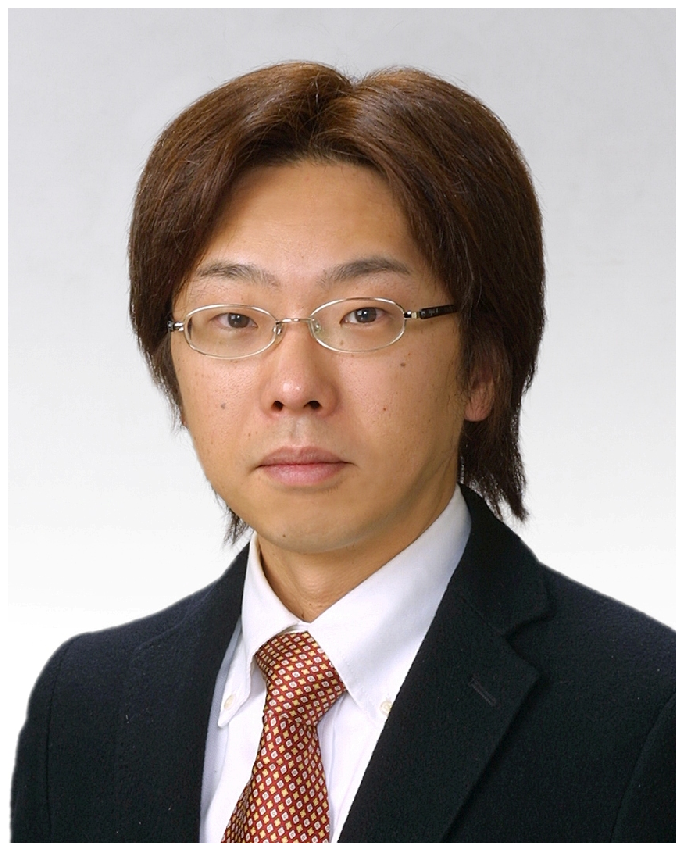}}]
  {Masaaki Nagahara}(S'00--M'03--SM'14)
  received the Bachelor's degree in engineering from Kobe University in 1998, 
the Master's degree and the Doctoral degree in informatics from Kyoto University 
in 2000 and 2003. He is currently a Senior Lecturer at Graduate School of Informatics, 
Kyoto University. His research interests include digital signal processing,
networked control, and sparse modeling. He received Young Authors Award in 1999
and Best Paper Award in 2012 from SICE,
Transition to Practice Award from IEEE Control Systems Society in 2012,
and Best Tutorial Paper Award from IEICE Communications Society in 2014.
He is a senior member of IEEE, and a member of SIAM, SICE, ISCIE and IEICE.
\end{IEEEbiography}
\begin{IEEEbiography}[
{\includegraphics[width=1in,height=1.25in,clip,keepaspectratio]{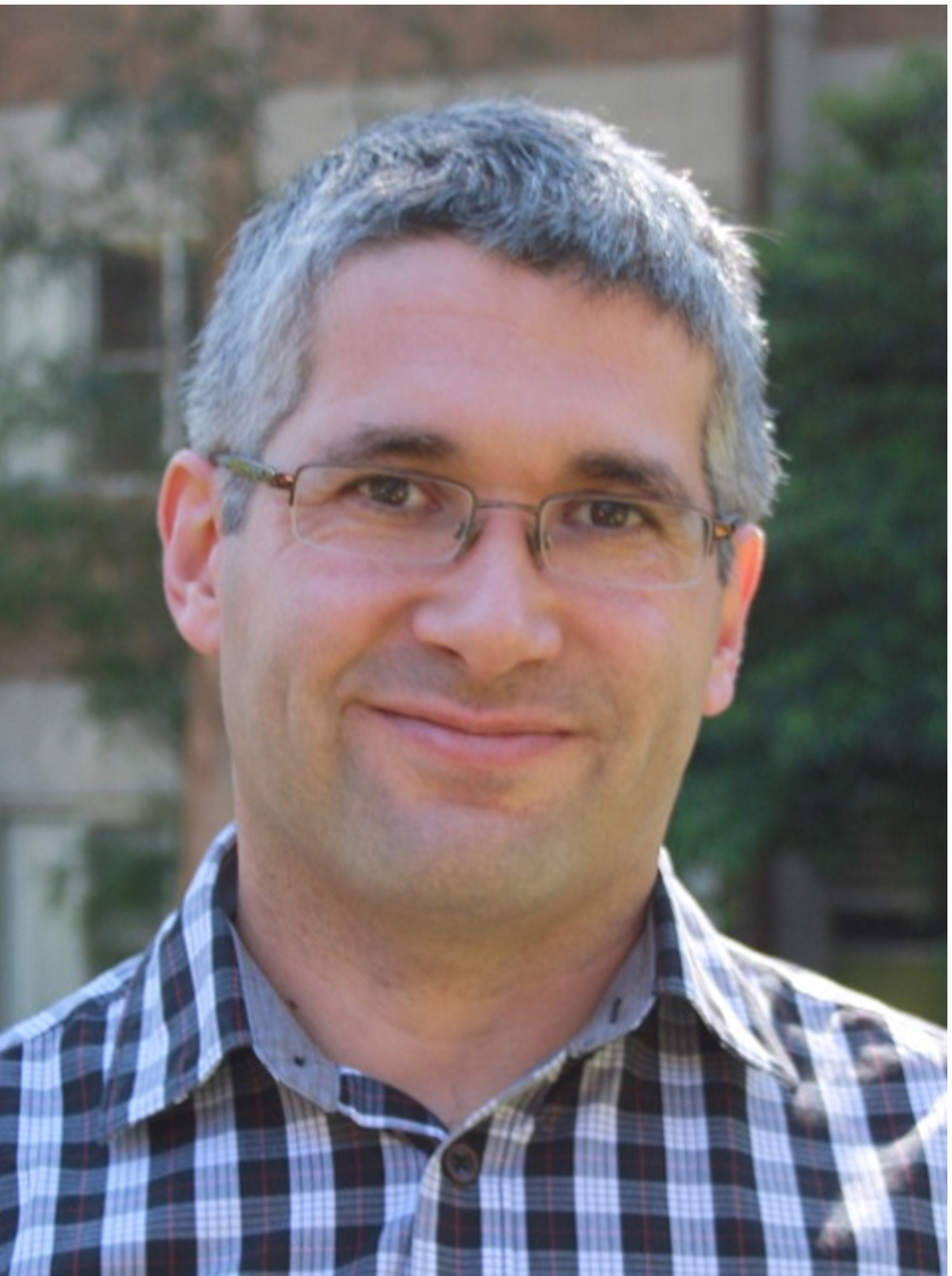}}]
  {Daniel E. Quevedo}(S'97--M'05--SM'14)
  holds the chair in Automatic
  Control (\emph{Regelungs- und Automatisierungstechnik})
  at the University of Paderborn,
  Germany. He received Ingeniero Civil Electr\'onico
  and M.Sc.\ degrees from the Universidad
  T\'ecnica Federico Santa Mar\'{\i}a,
  Chile, in 2000.  In 2005, he was awarded the Ph.D.\ degree
  from The University of Newcastle, Australia,
  where he subsequently  held various research academic positions. He has been a
  visiting researcher at various institutions, including Uppsala University,
  KTH Stockholm, Kyoto University,
   Karlsruher Institut für Technologie, 
   University of Notre Dame, INRIA Grenoble,
   The Hong Kong University of Science and
   Technology, Aalborg University, and NTU Singapore.  His  research
   interests include several areas within automatic control,
   signal processing, and power electronics. 

   \par Dr.\ Quevedo was supported by a full scholarship from the alumni
   association during his time at the Universidad 
   T\'ecnica Federico Santa Mar\'{\i}a and received several university-wide
   prizes upon graduating. He received the IEEE Conference on Decision and
   Control Best Student Paper Award in 2003 and was also a finalist  in 
   2002.  In 2009 he was awarded a five-year  Research Fellowship from
   the Australian Research Council.

   \par Prof.\ Quevedo is Editor  of the \emph{International Journal
     of Robust and Nonlinear Control} and
   serves as chair of the IEEE Control Systems Society
   \emph{Technical Committee on Networks \& Communication Systems}.
\end{IEEEbiography}
\begin{IEEEbiography}[
{\includegraphics[width=1in,height=1.25in,clip,keepaspectratio]{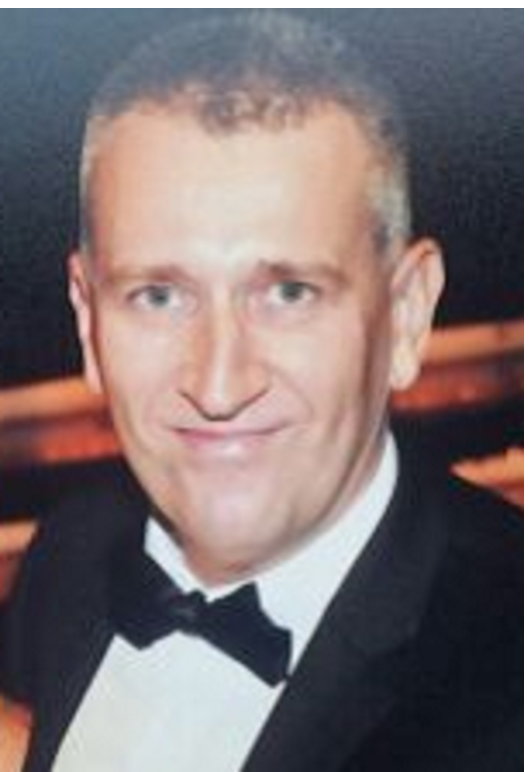}}]
{Dragan Ne\v{s}i\'{c}}(F'08)
is a Professor at the Department of Electrical and 
Electronic Engineering (DEEE) at The University of Melbourne, 
Australia. He received his B.E. degree in Mechanical Engineering 
from The University of Belgrade,Yugoslavia in 1990, and his Ph.D. 
degree from Systems Engineering, RSISE, Australian National University, 
Canberra, Australia in 1997. Since February 1999 he has been with 
The University of Melbourne. His research terests 
include networked control systems, discrete-time,sampled-data and 
continuous-time nonlinear control systems,input-to-state stability, 
extremum seeking control, applications of symbolic computation in control theory, 
hybrid control systems, and so on. He was awarded a Humboldt Research Fellowship 
(2003) by the Alexander von Humboldt Foundation, an Australian Professorial Fellowship
 (2004?2009) and Future Fellowship (2010?2014) by the Australian Research Council. 
He is a Fellow of IEEE and a Fellow of IEAust. He is currently a Distinguished Lecturer of CSS, 
IEEE (2008-). He served as an Associate Editor for the journals Automatica, IEEE Transactions
on Automatic Control, Systems and Control Letters and European Journal of Control.
\end{IEEEbiography}

\end{document}